# Large-Scale Education Reform in General Equilibrium: Regression Discontinuity Evidence from India

A Replication Study of Khanna (*Journal of Political Economy*, 2023)

David Roodman*

Open Philanthropy

**Abstract**

Mainly through regression discontinuity designs, Khanna (2023a) studies the impacts of a primary schooling expansion in India in the 1990s. Absent from the data set are four districts close to the modeled treatment discontinuity. Incorporating them cuts the impact of intention to treat on schooling attainment by two-thirds and the impact on wages by a third. Methodological revisions, including clustering by the geographic unit of treatment, double or triple standard errors, bringing estimates within a standard error of zero. These findings are robust to varying the location of the discontinuity, the bandwidth, and the radius of an exclusionary "donut." The estimates of the general equilibrium effect on the skill premium, as well as elasticities of substitution across age and skill groups, have high variance. One cause is that the treatment discontinuity does not occur quite where modeled. Moving the assumed cutoff produces sharper but wrong-signed results.

* david.roodman@openphilanthropy.org. Thanks to Gaurav Khanna for assistance with data and code and comments on an earlier draft, to anonymous referees for comments, and to Otis Reid for guidance. The packages esttab (Jann 2007), coefplot (Jann 2014), palettes (Jann 2022), blindschemes (Bischof 2017), grc1leg2 (Over 2022), and spmap (Pisati 2004) were used to make tables and graphs. Estimates were performed with rdrobust (Calonico et al. 2017) and parallel (Yon 2012). Data and code for this reanalysis are at [github.com/droodman/K23](https://github.com/droodman/K23). Included there is a response to comments from Khanna (2023b) on a previous version.

## Introduction

Khanna (2023a)—K23 for short—is the only study to estimate partial- and general-equilibrium effects of schooling in a developing country using a high-credibility research design. By exploiting a discontinuity in the eligibility of local governments for a primary school funding program in India, K23 estimates that the program increased future wages by 13.4% for each year of additional schooling, and would have increased them 6.6 percentage points more but for the general equilibrium effect of the expanded supply of skilled labor.

India's District Primary Education Programme (DPEP) received $1.5 billion from the World Bank and other donors in the mid-1990s (Jalan and Glinskaya 2013). It distributed the funds to districts—the second-level administrative units in India—to pay for school construction, teacher training, textbooks, and more. Some 271 of approximately 600 districts participated in the program, thereby increasing their primary education spending by an average of 17.5–20% for 5–7 years (Jalan and Glinskaya 2013, p. 4; Pandey 2000, p. 11). One criterion for funding was that a district's female literacy rate, as measured in the 1991 census, should be below the national average (Planning Commission 1994; Department of Education 1995). That average was 39.29%. As is often the case, treatment assignment did not perfectly follow the mathematical rule implied by official policy statements. Indeed, available documentation does not confirm that administrators imposed exactly 39.29%.[1] And districts could instead qualify by carrying out successful "Total Literacy Campaigns." K23 therefore estimates the impacts of intention to treat (ITT), defining it by the 39.29% cutoff. The main results are attained by applying regression discontinuity (RD) design to outcomes recorded in India's National Sample Survey of 2009–10, which was fielded when DPEP-affected children were young adults.

This comment assesses the robustness of the K23 findings by revising the data and methods. The changes are of three sorts. The data set is reconstructed from primary sources, which leads to some revisions. Techniques common in the analysis of survey data—clustering standard errors and incorporating inverse sampling probability weights—are incorporated (Deaton 1997). And the analysis is run according to published guidance for RD practice (Cattaneo, Idrobo, and Titiunik (CIT) 2019), which amounts to accepting defaults in popular software and systematically testing sensitivity to choices such as the cutoff and the bandwidth.

The empirical findings of K23 do not appear robust. See Table 1 for an overview. As shown in column 1, K23 estimates the impact of ITT on schooling attainment at 0.720 years. And ITT increases wages by 0.112 log points, for a return to schooling of 0.155 log points per year. In a separate and innovative calculation, K23 finds that the skill premium—the wage gain per extra year of schooling for more-skilled wage workers—was 0.065 log points *lower* in treated districts. Relative to the skill premium in untreated districts, this constitutes a 32% reduction.[2] K23 attributes this loss to a general equilibrium (GE) effect: DPEP stimulated the supply of skilled labor, lowering its price. A bootstrap-based 95% confidence interval for this impact estimate (not reported in K23) ranges between 3% and 77%. All these figures are included in the first column of the table as well.

[1] The official program guidelines refer only to "educationally backward districts with female literacy below the national average" (Department of Education 1995, p. 4).

The revisions to the data set substantially affect these results. Most important is the addition of four districts that are missing in the K23 data and that, though geographically dispersed, lie just above the 39.29% cutoff. Column 2 of the overview table shows that the revisions reduce the schooling impact estimate by 67% and the wage estimate by 35%. The point estimate for the return to schooling rises, but the standard error rises substantially more. Measured by the reported standard errors, these changes are surprisingly large. However, they become less surprising if the standard errors are themselves clustered by the geographic unit of treatment, as is conventional in this context. (See column 3.) With clustering, the GE effect is nearly unidentified.

In the last step of the overview, the robust, bias-corrected RD methodology of Calonico, Cattaneo, and Titiunik (CCT, 2014a) is fully applied. K23 uses the CCT algorithm to select the bandwidth in the schooling regressions, then transfers that bandwidth to the other regressions, and reports conventional RD point estimates and standard errors. The full CCT approach selects the bandwidth separately in each regression to (asymptotically) minimize its mean squared error. Trading off variance for bias in order to minimize MSE means allowing for some bias, which is caused by including observations farther from the discontinuity. Robust RD therefore performs a bias correction, and then adjusts the standard errors for the uncertainty of this correction. The robust RD results (final column) are also imprecise.

The particular instability of the GE estimates in the bottom panel of the table turns out to arise from the fuzziness of the treatment discontinuity at the ascribed location. The DPEP treatment rate falls sharply at 41% female literacy, not 39.29%. That weakens the 39.29%-based ITT dummy as an instrument for treatment. (Since the RD estimates in the upper panels are of the impacts of ITT., they do not involve the treatment variable and are more stable.) Perhaps DPEP's administrators conceived of the eligibility requirement as "40% or lower," with rates as high as 40.9% qualifying. Or perhaps some districts just above the cutoff were admitted according to a calculus that makes the observed cutoff endogenous to the outcomes of interest here. However, falsification tests, which look for discontinuities in pre-determined variables such as district area, provide no evidence of endogenous manipulation of the threshold. At any rate, because of this ambiguity, I report results under both thresholds. When using 41%, labor impacts are negative and the GE effect on the skill premium, though estimated more sharply, is also wrong-signed. It is hard to interpret those estimates as causal. Thus, the use of an alternative threshold does not resolve the lack of robustness in K23's empirical results.

[2] K23 (Table 3) reports 33% because of intermediate rounding.

Section 1 of this comment reconstructs the K23 data set. Section 2 revisits K23's preliminary graphical analysis. Section 3 does the same for the RD estimates. Section 4 discusses the program threshold and performs systematic robustness testing of the RD results. Section 5 brings the revisions to the estimation of general equilibrium effects. Section 6 concludes.

**Table 1. Results summary**

| | Original | + Revised data | + Cluster | + Robust, bias-corrected |
|---|---|---|---|---|
| *Impact of intention to treat on schooling: young (aged 17–34)* | | | | |
| Estimate | 0.720*** | 0.238 | 0.238 | 0.240 |
| | (0.199) | (0.201) | (0.476) | (0.505) |
| Bandwidth | 0.103 | 0.089 | 0.089 | 0.112 |
| Observations | 10,175 | 9,379 | 9,379 | 12,332 |
| *Impact of intention to treat on log wages: young* | | | | |
| Estimate | 0.112*** | 0.073** | 0.073 | –0.012 |
| | (0.031) | (0.032) | (0.106) | (0.125) |
| Bandwidth | 0.103 | 0.089 | 0.089 | 0.074 |
| Observations | 10,175 | 9,379 | 9,379 | 7,981 |
| *Impact of schooling on log wages: young* | | | | |
| Estimate | 0.155*** | 0.305 | 0.305 | 0.316 |
| | (0.043) | (0.224) | (0.564) | (0.589) |
| Bandwidth | 0.103 | 0.089 | 0.089 | 0.099 |
| Observations | 10,175 | 9,379 | 9,379 | 10,344 |
| *General equilibrium effect on skill premium* | | | | |
| Absolute impact (log points/year) | –0.065 | 0.703 | 0.703 | –0.188 |
| | [–0.19, –0.00] | [–8.76, 9.36] | [–2.79, 3.87] | [–21.0, 2.31] |
| Relative impact (%) | –32 | –29 | –29 | 213 |
| | [–77, –3] | [–86, 19] | [–492, 679] | [–3054, 3346] |

Notes: Except in the bottom panel, each cell reports results from a distinct regression. Estimates in the bottom panel are derived using the methodology of K23. Bandwidths are chosen with the Calonico, Cattaneo, and Titiunik (2014a) method except that in the final column the revised version from Calonico et al. (2019) is used. In all but the last column, bandwidths are optimized in the context of the regressions in the top row and copied to the rest. Results in column 1 exactly match K23 Table 1 and Table 3. Standard errors are in parentheses. In the bottom panel, percentile-based bootstrap percentile-based 95% confidence intervals are in brackets. *$p < 0.1$. **$p < 0.05$. ***$p < 0.01$.

## 1 Data set reconstruction

K23 studies impacts on a wide range of outcomes: school construction; schooling attainment, reading and math scores, migration, employment, and wage earnings; firms' adoption of skill-biased capital, local economic output, and private educational expenditures. I focus on the outcomes mentioned in the paper's abstract, introduction, and conclusion, which relate to schooling attainment and wage earnings.

From primary sources, I reconstruct the needed variables:

- *Female literacy rates* by district, the basis for ITT, are calculated from 1991 census data.[3]
- *Which districts participated in DPEP.* K23's source is not documented. I use a written answer from a minister to a question posed by a member of parliament in 2001. This source agrees with K23 that 271 districts participated in the program.[4]
- *Individual-level information on age, schooling attainment, and wage earnings from the 66th round of India's National Sample Survey (NSS).*[5] The survey was fielded in July 2009–June 2010, by which time children who were affected by DPEP were young adults.

The three data sets are linked by district, a process that is complicated by changes in district names and boundaries over time. Between 1991 and 2009, scores of districts subdivided, producing new, "single-parent districts." Others were formed from fragments of several antecedents, producing "multi-parent districts." I track the evolution from 1991 to 2001 using Kumar and Somanathan (2016, Tables 7, 8, 9d) and from 2001 to 2009 using Wikipedia and official reports from the 2011 census.[6] Like K23, I retain single-parent districts since they do not threaten the consistency of the K23 estimators: if a parent district qualified for intention to treat by virtue of low female literacy, we may class its subparts the same way. *Multi*-parent districts pose a problem, since some descend from districts with significantly different female literacy rates. I exclude any district of this type unless the parents had nearly the same female literacy rate or else all but one parent contributed *de minimus* population.[7]

Table 2 and the top two-thirds of Figure 1 compare the original and new ITT and treatment variables, with reference to the district boundaries of 2009. The two versions of ITT agree except that the K23 version is marked missing in 15 districts for which the new one is not, and the new one is marked missing in 13 districts (because of multiple parents) where the K23 one is not.[8] Appendix A provides more detail on these disagreements.

---

[3] District-level primary census abstracts for 1991 were downloaded from censusindia.gov.in/census.website/data/census-tables.

[4] datais.info/loksabha/question/db0cac20ad912c779f1de1c7b7fd60f3/DISTRICT+PRIMARY+EDUCATION+PROGRAMME.

[5] icssrdataservice.in/datarepository/index.php/catalog/89/data_dictionary.

[6] I use the census reports allow to estimate the population contributions of parent districts to child districts.

[7] For each 2009 district, I compute the standard deviation of the parents' literacy rates, weighting by their population contributions to the child. Multi-parent districts are retained only if the standard deviation falls below 1%. The threshold was 3% in an earlier version of this paper. I lowered it to 1% after Gaurav Khanna pointed out that the 3% threshold happened to admit two multi-parent districts, Champawat and Moga, that received non-*de minimus* population contributions from districts on either side of the female literacy cutoff. I thank Khanna for the critique.

The outcome variables taken from the 2009–10 NSS differ slightly in construction. The new schooling attainment variable is 0 years rather than 1 or 2 for individuals self-reporting as "literate without formal schooling." Some 2,000 of 72,000 respondents in the sample report wage earnings from multiple work activities; in the new data, wages are summed from all activities rather than the first-listed one. On overlap, the original and new age variable match exactly, the female literacy and years of schooling variables are correlated 0.9998, and the log wage variables are correlated 0.9968.

More important is a revision in sample. Four districts are wholly absent in the K23 data file but present with complete observations in the new one. They can be discerned, in grey, in the bottom-left of Figure 1. Though geographically dispersed, they are statistically close in one sense. All had female literacy rates just above the ascribed DPEP eligibility threshold of 39.29%: Aurangabad district in Maharashtra state (at 39.64%), Tamenglong in Manipur (39.68%), Cuddalore in Tamil Nadu (39.70%), and Latur in Maharashtra (39.74%).[9] All but Tamenglong received DPEP funding, so three of the four are non-compliers under an ITT rule based on the 39.29% threshold.

**Table 2. Cross-tabulation of original and new intention to treat and treatment variables**

Intention to treat (female literacy in 1991 below 39.29%)

| | Original | | | |
|---|---|---|---|---|
| New | 0 | 1 | Missing | Total |
| 0 | 231 | 0 | 7 | 238 |
| 1 | 0 | 327 | 8 | 335 |
| Missing | 6 | 7 | 32 | 45 |
| Total | 237 | 334 | 47 | 618 |

Treatment (received DPEP funding)

| | Original | | | |
|---|---|---|---|---|
| New | 0 | 1 | Missing | Total |
| 0 | 303 | 4 | 24 | 331 |
| 1 | 45 | 219 | 16 | 280 |
| Missing | 0 | 0 | 7 | 7 |
| Total | 348 | 223 | 47 | 618 |

Note: Unit of observation is the 2009 district.

[8] Because of district subdivisions, more than 271 districts-of-2009 are classed as treated in the revised data set.

[9] The bottom left of Figure 1 shows *five* districts absent in the original and present in the new data set. The fifth (and westernmost), Gandhinagar, is also excluded from the new analysis because it is multi-parent. And the population-weighted average 1991 female literacy rate of its parents is 65%, which is far from the identifying cutoff.

**Figure 1. Overview of original and new data sets**

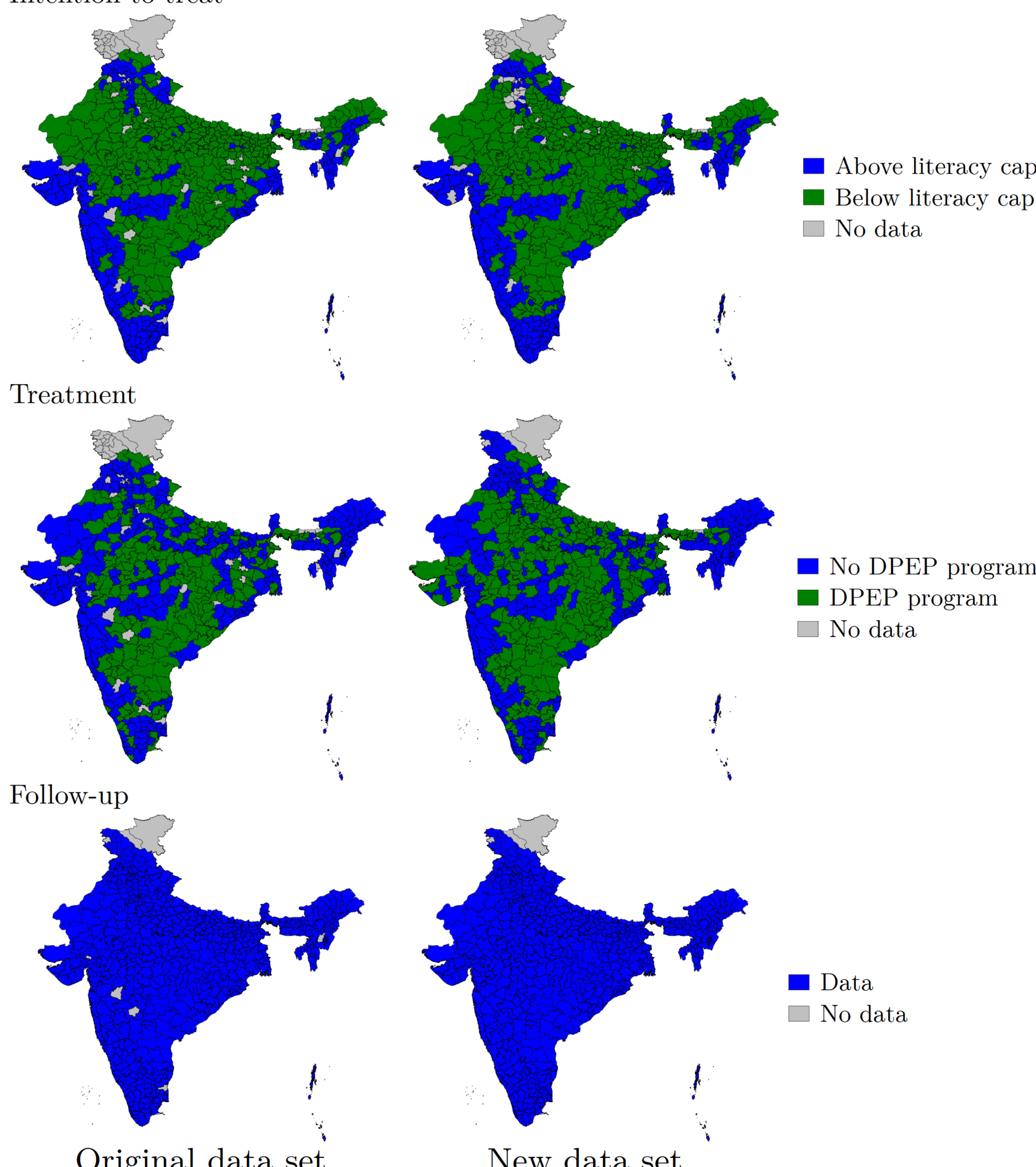

Notes: Original data set is from K23. Sources for the new data set are described in text. “Intention to treat” is whether a district’s female literacy rate in 1991 was below the national average of 39.29%. Treatment is whether a district participated in the District Primary Education Programme. Follow-up is whether the district appears in the extract from the National Sample Survey, 66$^{th}$ round, 2009–10. Underlying map from Minnesota Population Center (2020).

## 2 Graphical preliminaries

In their guide to RD practice, Imbens and Lemieux (2008) recommend checking graphically for discontinuities in the dependent variables of interest. "If the basic plot does not show any evidence of a discontinuity, there is relatively little chance that the more sophisticated analyses will lead to robust and credible estimates with statistically and substantially significant magnitudes." In that spirit, K23 begins the empirics with graphs, most made with the rdplot program for Stata (CCT 2015). The program plots scatters of a dependent variable against a running variable after averaging the first within bins of the second. By default, the bins are chosen by a data-driven algorithm and cover the full range of the running variable. The program then overlays polynomials fits, quartic by default, to the un-binned data on either side of the threshold. Users may override the defaults governing the polynomial order, the binning, and the sample. In K23, the fits are quadratic rather than quartic. The bin counts are hand-coded: for DPEP assignment, 18 on the left and 22 on the right; for years of schooling, 6 and 7; for log wages, 8 and 8. For the latter two variables, the sample is restricted to within 0.2 of the cutoff. No motivation for these choices is recorded.

Figure 2 below shows the effects of reverting to the rdplot defaults and switching to the new data set. Plots in the first column are exact replications of the original. They give the impression of discontinuities with the predicted sign in treatment, schooling, and wages. The second column accepts rdplot's defaults. These changes preserve the appearance of a discontinuity in treatment near 39.29% female literacy (top row). They do not definitively alter the picture for schooling. (Note the expanded vertical scale.) They do remove the impression of a discontinuity in wages.

The final column of Figure 2 switches to the new data set. This removes the discontinuity in schooling too. And the influence of the four previously missing districts is discernible in the treatment plot, just to the right of the cutoff (upper right plot in the figure). The smoothed fit now slopes more steeply there, probably because three of the four are treated, making them non-compliers. This does not eliminate the apparent discontinuity in treatment, but does reduce it.

To the extent that the K23 plots support an inference of significant program impacts, it is worth bearing in mind that this impression is correlated with overrides of software defaults and the absence of some districts near the threshold.

**Figure 2. Global polynomial fits of program assignment, schooling attainment, and log wages to female literacy**

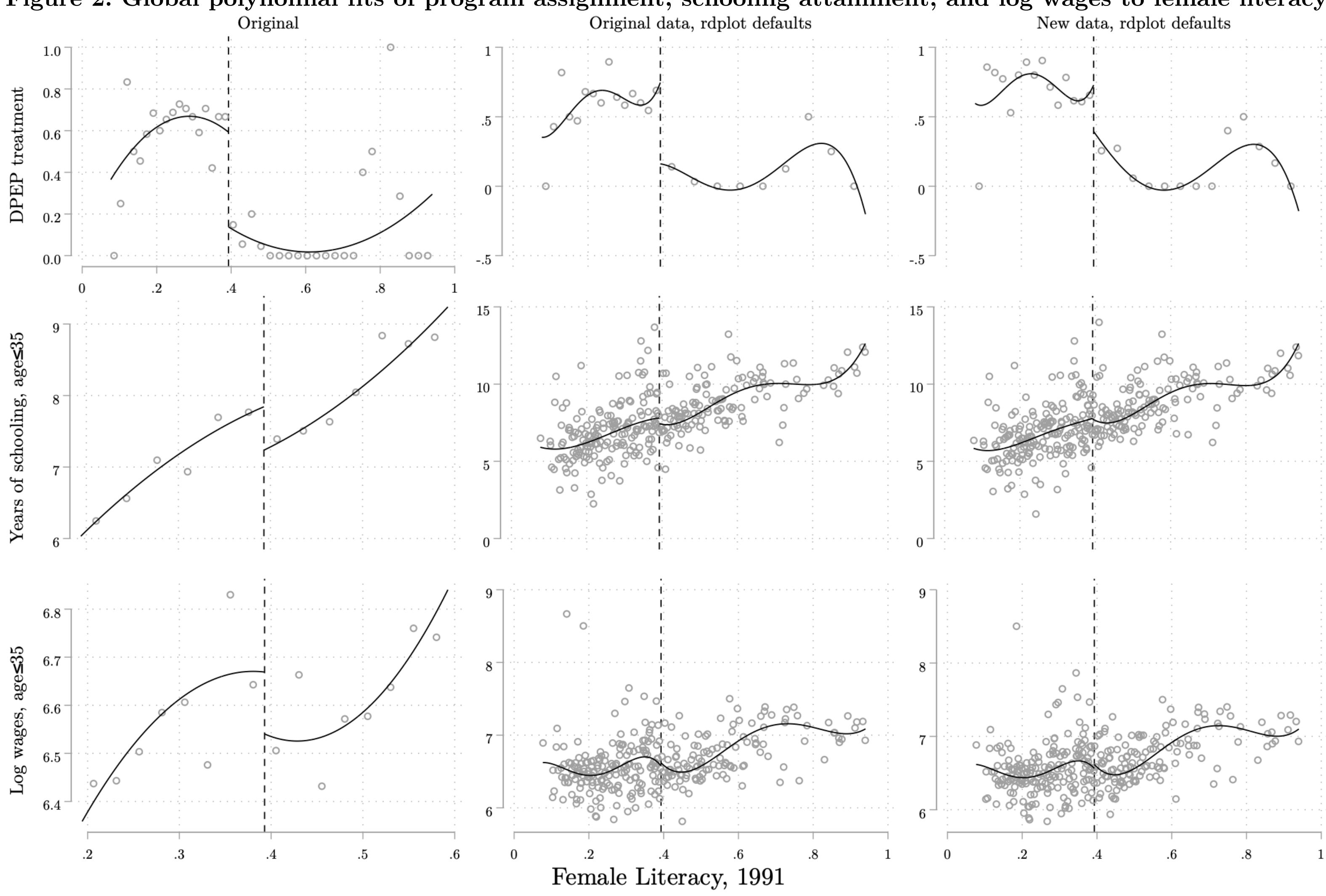

Notes: Each plot is made by the rdplot program for Stata (Calonico et al. 2017). Observations are averaged within bins before scatter-plotting. Polynomials are fit to the un-binned data on each side of the 39.29% female literacy threshold. Plots in the first column exactly match ones in Khanna (2023a, Figures 1 and 2). They entail overriding rdplot defaults for the order of the polynomial fit and the number of bins, and, in rows 2 and 3, restricting the sample to within 0.2 points of the cutoff. Plots in the second column are made by accepting rdplot's defaults. Plots in the third column additionally switch to the new data set.

# 3 Regression discontinuity estimates

The main K23 RD regressions estimate the impact of ITT on the treated—living in a district that score just below the 39.29% female literacy threshold affected schooling attainment and wages. Fuzzy regression discontinuity (FRD) regressions essentially take ratios of the RD estimates to calculate the returns to schooling. In an appendix, K23 also uses FRD to estimate the impact of treatment on the treated. This section focuses on the ITT estimates in the main text. It incorporates the data changes described in section 1 and revises the K23 methodology in certain respects.

## 3.1 Robust RD

K23 performs RD and FRD with the rdrobust program for Stata (CCT 2014b; Calonico et al. 2017), which implements the "robust RD" method of CCT (2014a), as revised in Calonico, Cattaneo, and Farrell (2018) and Calonico et al. (2019). By default, the CCT procedure runs as follows:

1. Given an RD specification, a bandwidth is chosen to define the radius of the sample. The selection algorithm is designed to balance bias and variance in order to asymptotically minimize mean squared error. (Raising the bandwidth would increase sample size, adding precision. But it could add bias by giving more weight to observations farther from the threshold.) Many aspects of the specification affect the optimal bandwidth: the running variable; the threshold; the response variable; the order of the polynomial model on each side of the cutoff; the weighting kernel type, such as uniform or triangular; the choice of variance estimator, such as clustered; and the observation weights, if any.
2. The impact is estimated by fitting the model to the data within the bandwidth, separately on each side of the threshold. The polynomials for these local regressions are linear—again, by default—in contrast with the higher-order global polynomial model in the graphical preliminaries. The weighting kernel is triangular.
3. Since the bias in RD results from curvature in the regression function near the cutoff, steps 1 and 2 are repeated with a quadratic model to estimate the local curvature and generate a bias estimate. The latter is subtracted from the initial point estimate.
4. Like all such bias corrections, this one contains uncertainty and so increases variance. "Robust" standard errors are computed to reflect the additional variance.

Simulations in CCT (2014a, Table 1) show that robust RD performs better than conventional RD with the CCT bandwidth and sometimes much better than conventional RD with the bandwidth selector of Imbens and Kalyanaraman (IK, 2012)—where "performs better" means less over-rejection of the null.

K23 departs from this sequence in two ways. First, it stops at step 2. It runs conventional RD with the CCT or IK bandwidth.[10] Second, K23 overrides the default behavior of the estimation software in order to use the same bandwidth for all regressions—that is, for all outcomes (years of education, log earnings, return to schooling) and all subgroups (young and old). Because the MSE-optimized bandwidths of nearly all the other regressions are smaller, as will be shown, this decision amplifies the concern about bias from higher bandwidths. For these reasons, I will switch to robust RD, following all four steps above.

### 3.2 Clustering

K23 does not cluster the standard errors in the main RD regressions. But arguments for clustering in regressions on microdata when the treatment is assigned at the geographic level do apply (Bertrand, Duflo, and Mullainathan 2004; Bartalotti and Brummet 2017; Calonico et al. 2019; CIT 2019, p. 74).[11] Abadie et al. (2023) additionally justifies a clustering correction when, as here, the complex survey design disproportionately samples some districts. Other DPEP impact studies cluster by district (Azam and Saing 2017; Sunder 2018; Kuipers, Nellis, and Stommes 2025). In fact, K23 clusters in its DID regressions and does report clustered RD results in an appendix.

But K23's clustered results come with some limitations. The clustering is not incorporated into bandwidth selection. Only schooling impacts are reported. And in one of the methods used—making "nearest neighbor" comparisons to estimate residuals (Calonico et al. 2019)—the standard error can range from about 0.1 to 0.4 depending on the number of nearest neighbors chosen. K23 uses 48 nearest neighbors, for reasons that are not stated, yielding a standard error of about 0.2 for schooling impacts. In the non-clustered, heteroskedasticity-only context, simulations in Calonico et al. (2019, Tables SA-1 and SA-2) show nearly identical performance from the usual residual plug-in method and the nearest-neighbor method. No simulation evidence is available to rebut the presumption that the methods would perform similarly in a clustered context. And there is no guidance on setting the number of nearest neighbors when clustering. For these reasons, it seems best to cluster with the familiar plug-in method. Since the district of ITT assignment (based on a 1991 district's female literacy rate) is a somewhat coarser unit that the district of follow-up (as of 2009), I will cluster by the former. This should adjust for error dependence created by clustering of treatment and clustering of follow-up sampling.

[10] CIT (2019, p. 59) criticizes such an approach: "It would be intellectually and methodologically incoherent to simultaneously select a bandwidth according to a bias-variance trade-off and then proceed as if the bias were zero, that is, as if the local polynomial fit were exact and no misspecification error existed." CCT (2014a, p. 2295) explains that the MSE-minimizing bandwidth is "too 'large' to ensure the validity of the distributional approximations usually invoked." As a result, it is "by construction not valid for conventional confidence intervals....This implies that conventional confidence intervals may substantially over-reject the null hypothesis of no treatment effect."

[11] Since the running variable, female literacy, is continuous and unique for each 1991 district, I cluster by the running variable as an expedient to clustering by 1991 district. Kolesár and Rothe (2018) warns against clustering by the running variable when it is discrete. The warning does not apply here because the running variable takes some 200 values within a typical K23 bandwidth of 0.1, and because clustering here increases standard errors rather than decreasing them as in the degenerate example of Kolesár and Rothe.

### 3.3 Weighting

K23 does not incorporate survey weights. The NSS sampling program is presumptively endogenous to the outcomes studied in K23. In the second sampling stage, households are stratified by assets and economic activity (in rural areas) or per-capital expenditure (in urban areas).[12] Not factoring in the inverse-sampling-probability weights might make the estimators inconsistent (Hausman and Wise 1981; Solon, Haider, and Wooldridge 2015; Roodman [2026]). On the other hand, the distribution of the weights is so skewed—with a median of 895 and the 99th percentile at 18,757—that incorporating them can greatly increase the variance of estimators. To limit the skew, I will prefer a procedure simulated in Potter and Zheng (2015), which is to winsorize high values to the median weight plus 4 times the interquartile range.

### 3.4 Results

The data comments in section 1 and the methodological comments above lead to revisions of the K23 RD analysis. I view most of the changes as non-controversial, at least as minimally arbitrary robustness tests. It makes sense to include all districts with complete data. The changes to variable construction are reasonable and minor. The arguments and precedents for clustering by district are strong. The robust RD methodology literally goes by the book (CIT 2019) and amounts to taking defaults in the dominant RD software package. The addition of weights, however, may pose a substantial tradeoff between bias and variance. I therefore replicate key K23 estimates and then incorporate the revisions cumulatively, saving weighting for last.

The results appear in Table 3. There are panels for the RD schooling and log wages regressions as well as for the FRD returns-to-schooling regressions. As in K23, the impacts are estimated separately for young and old, i.e., those under 35 and those over. The first two columns replicate most of K23 Table 1. Column 1 uses the CCT bandwidth selector and column 2 the IK selector. The CCT-based results suggest that living in a district just under the female literacy threshold lifted schooling attainment among under-35s by 0.72 years (standard error 0.199) and wage earnings by 0.112 (0.031) log points. The estimate of the return to schooling is 0.155 (0.043) log points per year. Older people (over-35s) were born too late to benefit directly from DPEP and, indeed, the impact estimates for them are much closer to zero. Notice that the bandwidths are the same within each of these columns, as they are derived as optimal for the first regression and copied to the rest.

[12] microdata.gov.in/nada43/index.php/catalog/124/sampling

In the columns 3 and 4, variables are revised as described in section 1. The results hardly change.

In columns 5 and 6, the four missing districts are also added. This dramatically affects results. Using the CCT bandwidth, the schooling impact among the young changes from 0.699 (0.178) years to 0.238 (0.201), the wage impact from 0.136 (0.028) log points to 0.073 (0.034), and the returns to schooling from 0.195 (0.046) to 0.305 (0.224). The numbers are similar with the IK bandwidth.

Clustering is introduced next, in columns 7 and 8. For clarity, it is not yet factored into the bandwidth selection. As result, the bandwidths and point estimates do not change, even as the standard errors double or triple. All estimates are now less than a standard error from zero.

Starting in column 9, the estimates are performed with regression-specific optimized bandwidths. As well, clustering is now fully incorporated into the bandwidth selection step, which requires switching from the 2014 to the 2016 version of rdrobust (Calonico et al. 2019). That in turn means that the IK bandwidth selection algorithm is no longer available, so only CCT-based estimates are reported.[13] (The next section will assess sensitivity to bandwidth more thoroughly.) The bandwidths, point estimates, and standard errors do not change much.

Next, CCT's bias correction is applied, which preserves standard errors but shifts point estimates—in this case, mostly downward. Then, in column 11, CCT's robust standard errors are reported, which adjust for the uncertainty added by the bias correction. Here, the point estimates hold fixed while the standard errors widen. And while the narrative took many steps to reach this specification, it is worth noting that the rdrobust package runs it largely by default. (Only the clustering needs to be specified by the user.)

Finally, survey weights are incorporated. Perhaps because the extreme weights are winsorized, the change only increases variance by about 10%. Point estimates shift with no apparent pattern and little statistical significance.

The numerical results confirm the impression from the graphical analysis. The main K23 RD results are not robust to changes that ought to bring improvements in bias, consistency, and inference.

[13] Calonico et al. (2017, p. 381) present the bandwidth selectors in the 2016 version as upgrades from the CCT, IK, and cross-validation options in the 2014 version.

**Table 3. Replication and revision of Khanna (2023a) RD regressions**

| | (1) | (2) | (3) | (4) | (5) | (6) | (7) | (8) | (9) | (10) | (11) | (12) |
|---|---|---|---|---|---|---|---|---|---|---|---|---|
| *Impact of intention to treat on schooling: young (sharp RD)* | | | | | | | | | | | | |
| Estimate | 0.720*** | 0.698*** | 0.699*** | 0.743*** | 0.238 | 0.308 | 0.238 | 0.308 | 0.349 | 0.240 | 0.240 | 0.143 |
| | (0.199) | (0.173) | (0.178) | (0.194) | (0.201) | (0.189) | (0.476) | (0.452) | (0.434) | (0.434) | (0.505) | (0.554) |
| Bandwidth | 0.103 | 0.135 | 0.124 | 0.107 | 0.089 | 0.103 | 0.089 | 0.103 | 0.112 | 0.112 | 0.112 | 0.140 |
| Observations | 10,175 | 14,277 | 13,669 | 11,034 | 9,379 | 10,485 | 9,379 | 10,485 | 12,332 | 12,332 | 12,332 | 14,946 |
| *Impact of intention to treat on log wages: young (sharp RD)* | | | | | | | | | | | | |
| Estimate | 0.112*** | 0.145*** | 0.136*** | 0.123*** | 0.073** | 0.098*** | 0.073 | 0.098 | 0.030 | –0.012 | –0.012 | 0.061 |
| | (0.031) | (0.027) | (0.028) | (0.031) | (0.032) | (0.030) | (0.106) | (0.101) | (0.110) | (0.110) | (0.125) | (0.140) |
| Bandwidth | 0.103 | 0.135 | 0.124 | 0.107 | 0.089 | 0.103 | 0.089 | 0.103 | 0.074 | 0.074 | 0.074 | 0.077 |
| Observations | 10,175 | 14,277 | 13,669 | 11,034 | 9,379 | 10,485 | 9,379 | 10,485 | 7,981 | 7,981 | 7,981 | 8,256 |
| *Impact of schooling on log wages: young (fuzzy RD)* | | | | | | | | | | | | |
| Estimate | 0.155*** | 0.208*** | 0.195*** | 0.165*** | 0.305 | 0.318* | 0.305 | 0.318 | 0.333 | 0.316 | 0.316 | 0.439 |
| | (0.043) | (0.046) | (0.046) | (0.042) | (0.224) | (0.169) | (0.564) | (0.424) | (0.499) | (0.499) | (0.589) | (0.694) |
| Bandwidth | 0.103 | 0.135 | 0.124 | 0.107 | 0.089 | 0.103 | 0.089 | 0.103 | 0.099 | 0.099 | 0.099 | 0.108 |
| Observations | 10,175 | 14,277 | 13,669 | 11,034 | 9,379 | 10,485 | 9,379 | 10,485 | 10,344 | 10,344 | 10,344 | 11,352 |
| *Impact of intention to treat on schooling: old (sharp RD)* | | | | | | | | | | | | |
| Estimate | –0.086 | 0.100 | –0.031 | –0.075 | –0.153 | –0.178 | –0.153 | –0.178 | –0.138 | –0.425 | –0.425 | 0.565 |
| | (0.218) | (0.188) | (0.194) | (0.213) | (0.223) | (0.209) | (0.679) | (0.678) | (0.654) | (0.654) | (0.721) | (0.859) |
| Bandwidth | 0.103 | 0.135 | 0.124 | 0.107 | 0.089 | 0.103 | 0.089 | 0.103 | 0.125 | 0.125 | 0.125 | 0.097 |
| Observations | 11,293 | 16,007 | 15,293 | 12,376 | 10,397 | 11,730 | 10,397 | 11,730 | 15,765 | 15,765 | 15,765 | 11,319 |
| *Impact of intention to treat on log wages: old (sharp RD)* | | | | | | | | | | | | |
| Estimate | –0.011 | 0.043 | 0.012 | –0.009 | 0.001 | 0.012 | 0.001 | 0.012 | –0.004 | –0.052 | –0.052 | 0.122 |
| | (0.037) | (0.032) | (0.033) | (0.037) | (0.038) | (0.036) | (0.145) | (0.139) | (0.147) | (0.147) | (0.168) | (0.157) |
| Bandwidth | 0.103 | 0.135 | 0.124 | 0.107 | 0.089 | 0.103 | 0.089 | 0.103 | 0.085 | 0.085 | 0.085 | 0.113 |
| Observations | 11,290 | 16,004 | 15,293 | 12,376 | 10,397 | 11,730 | 10,397 | 11,730 | 9,971 | 9,971 | 9,971 | 13,905 |
| *Impact of schooling on log wages: old (fuzzy RD)* | | | | | | | | | | | | |
| Estimate | 0.129 | 0.442 | –0.381 | 0.117 | –0.009 | –0.070 | –0.009 | –0.070 | –0.059 | 0.367 | 0.367 | 0.563 |
| | (0.303) | (0.666) | (3.242) | (0.351) | (0.258) | (0.265) | (0.977) | (1.011) | (0.887) | (0.887) | (1.009) | (0.795) |
| Bandwidth | 0.103 | 0.135 | 0.124 | 0.107 | 0.089 | 0.103 | 0.089 | 0.103 | 0.110 | 0.110 | 0.110 | 0.143 |
| Observations | 11,290 | 16,004 | 15,293 | 12,376 | 10,397 | 11,730 | 10,397 | 11,730 | 13,247 | 13,247 | 13,247 | 17,183 |
| Bandwidth method | CCT | IK | CCT | IK | CCT | IK | CCT | IK | CCT | CCT | CCT | CCT |
| Revised variables | | | ✓ | ✓ | ✓ | ✓ | ✓ | ✓ | ✓ | ✓ | ✓ | ✓ |
| Four missing districts | | | | | ✓ | ✓ | ✓ | ✓ | ✓ | ✓ | ✓ | ✓ |
| Clustered | | | | | | | ✓ | ✓ | ✓ | ✓ | ✓ | ✓ |
| Regression-specific BW | | | | | | | | | ✓ | ✓ | ✓ | ✓ |
| Bias-corrected | | | | | | | | | | ✓ | ✓ | ✓ |
| Robust RD | | | | | | | | | | | ✓ | ✓ |
| Weights | | | | | | | | | | | | ✓ |

Notes: Each cell reports results from a distinct RD regression—sharp or fuzzy, as indicated. CCT indicates the Calonico, Cattaneo, and Titiunik (2014a) bandwidth selection method. IK indicates the Imbens and Kalyanaraman (2012) method. Results in columns 1 and 2 exactly match K23 Table 1. Successive columns introduce changes to data and specification, cumulatively. Starting in column 9, clustering is factored into the bandwidth selection using the 2016 version of the rdrobust package, which does not offer the IK method. In the final column, weights are winsorized to the median weight plus 4 times the interquartile range. "Young" means aged 17–34 at follow-up and "old" means 36–75. Standard errors are in parentheses. *$p < 0.1$. **$p < 0.05$. ***$p < 0.01$.

## 4 Systematic robustness testing

Just as graphical preliminaries are standard before RD estimation, sensitivity testing is standard after. It too can be carried out graphically. Here, I check for sensitivity to moving the identifying cutoff, varying the bandwidth around it, and creating a "donut" hole to drop districts closest to the cutoff. I focus on the impacts of ITT on the young in the preferred specification reached in the last column of Table 3. Since K23 does not test robustness in this particular manner, this section makes a temporary departure from the narrow path of replication and revision. The analysis here arose through interaction with the original author (Khanna 2023b).

### 4.1 Varying the intention-to-treat cutoff

Although official documents make clear that DPEP was targeted at districts with female literacy below the national average (Planning Commission 1994, §17.176; Department of Education 1995, p. 4), I have found no precise statement of the threshold used in practice. We can compute that 129,752,482 of 330,286,606 Indian females over age 6—39.29%—were marked as literate in the 1991 census. It does not follow that the central government cleaved the field at precisely that rate. Plausibly, officials implemented a "39% or less" rule in such a way that even 39.99% qualified. That would have put the exact cutoff at 40.00%. Or perhaps in an administrative game of telephone, or in the face of the difficulty of rejecting districts that almost qualified, "below 40%" in practice became "40% or below." That would have lifted the effective threshold to 41.00%. If the true and modeled thresholds differ, this will add noise to the ITT indicator and reduce power.

Figure 3 examines the question of threshold placement by plotting the fraction of districts with female literacy in $[0.10, 0.11)$, $[0.11, 0.12)$, etc., that received DPEP funding. The threshold most consistent with the data is 0.41. The treatment rate does not decline as one approaches 0.41 from the left. But it plunges to 0 for the 22 districts between 0.41 and 0.43, returns to a lower level between 0.43 and 0.49, then falls back to zero. All of the DPEP districts with literacy near 0.8 are in Kerala; they qualified on the alterative criterion of having run successful Total Literacy Campaigns (Pandey 2000, p. 15). Some of the districts between 0.41 and 0.43 may have qualified in the same way. Since it is plausible on priors that the threshold in practice was a whole percentage, Bayesian reasoning puts nontrivial weight on 0.40 and 0.41 as the values closest to the administrative reality.

If using a different threshold would make RD estimation more powerful, it might add bias as well. Officials may have tuned the administrative threshold in response to traits of marginal districts, such as poverty or political power. That would undermine the local exogeneity of ITT. To examine this concern, I follow guidance from CIT (2019, p. 79), performing a placebo check for discontinuities in several predetermined, district-level variables at the candidate thresholds of 0.3929, 0.40, and 0.41.

In particular, I check for discontinuities in district population, area, number of residential houses, number of households, and the share of those over age 6 engaged in agricultural work; all come from the same district-level primary census abstracts as the female literacy rates.[14] See Table 4. The unit of observation is the 1991 district. Under the null that ITT is locally exogenous at each candidate threshold, districts just on either side will not differ systematically on any traits. The high $p$ values from the robust RD regression reported in Table 4 do not reject that null.

Having established the plausibility of alternative cutoffs, I examine how the RD estimates of impacts on the young evolve as the cutoff varies across a range. See Figure 4. The top pane tests for evidence of an impact of intention to treat on treatment. Interestingly, with the weighting, unlike in Figure 2, there is no longer an apparent jumped at the asserted cutoff, 0.3929. But, as expected, the a break does appear as one shifts the cutoff toward 0.41. The lower three panes of Figure 4 show that the previous section's finding of no clear, positive impact on schooling or wages is robust to varying the cutoff. The only major update is that using 0.41 leads to a negative wage impact estimate (third pane). Appendix B reruns the full RD analysis from previous section using the 0.41 cutoff. The estimated impact on the wages of the young is strongly negative in nearly all specifications.

### 4.2 Varying the bandwidth

In these regressions, the MSE-minimizing bandwidth is estimated with error. The CCT algorithm is only guaranteed to work if certain assumptions are fully met, and even then only asymptotically. So I next check for sensitivity in the preferred specifications to varying the bandwidth while fixing the female literacy cutoff at 0.3929 or 0.41 (CIT, p. 94).[15] The results are in Figure 5. The vertical lines now show the bandwidths that would be selected by the CCT algorithm. The impact estimates are generally stable across bandwidths, which is a sign that the CCT bias correction is at least partially compensating for the rise in bias as the bandwidth increases. When using the 0.41 threshold, moving away from the CCT bandwidth does make the apparent wage impacts less negative.

### 4.3 Varying the radius of a donut hole

A common practice in RD is to drop the observations closest to the cutoff. Historically, the technique was developed in response to the concern that agents could manipulate a running variable, such as recorded birthweight, in order to land on the desirable side of a cutoff (Almond and Doyle

[14] The census abstracts decompose population by age only to the extent of reporting 6-and-under totals, separately for males and females.

[15] Ideally, these tests would set the bandwidth for the estimation step while allowing rdrobust to choose the bandwidth for the bias-correction step through a data-driven algorithm. However, if one sets the estimation bandwidth to a specific value, rdrobust assigns the same value to the bias-correction bandwidth. To increase realism, I therefore manually assign both. I make the ratio between the two the same as when both are chosen automatically by the CCT MSE-minimizing algorithm.

2011; Barreca, Lindo, and Waddell 2016; Catteneo, Idrobo, and Titiunik 2019). That concern does not apply here. Census respondents in 1991 did not manipulate their answers in anticipation of a donor-driven education finance program not yet formulated. Even if they foresaw the program, they could not have known what national average would emerge from the census tabulation on female literacy. Regardless, even when manipulation is unlikely, CIT (p. 92) favors donut RD as a sensitivity test. Results of varying the radius of the hole from 0% to 1% are exhibited in Figure 6. The findings are once more stable, except that with the cutoff at 39.29%, the apparent impact on schooling temporarily rises into statistical significance when the radius is about 0.4%.[16]

Overall, the preferred RD estimates in the previous section do not appear fragile, except in that a higher cutoff leads to a negative wage impact estimate.

**Figure 3. Fraction of 2009 districts receiving DPEP treatment**

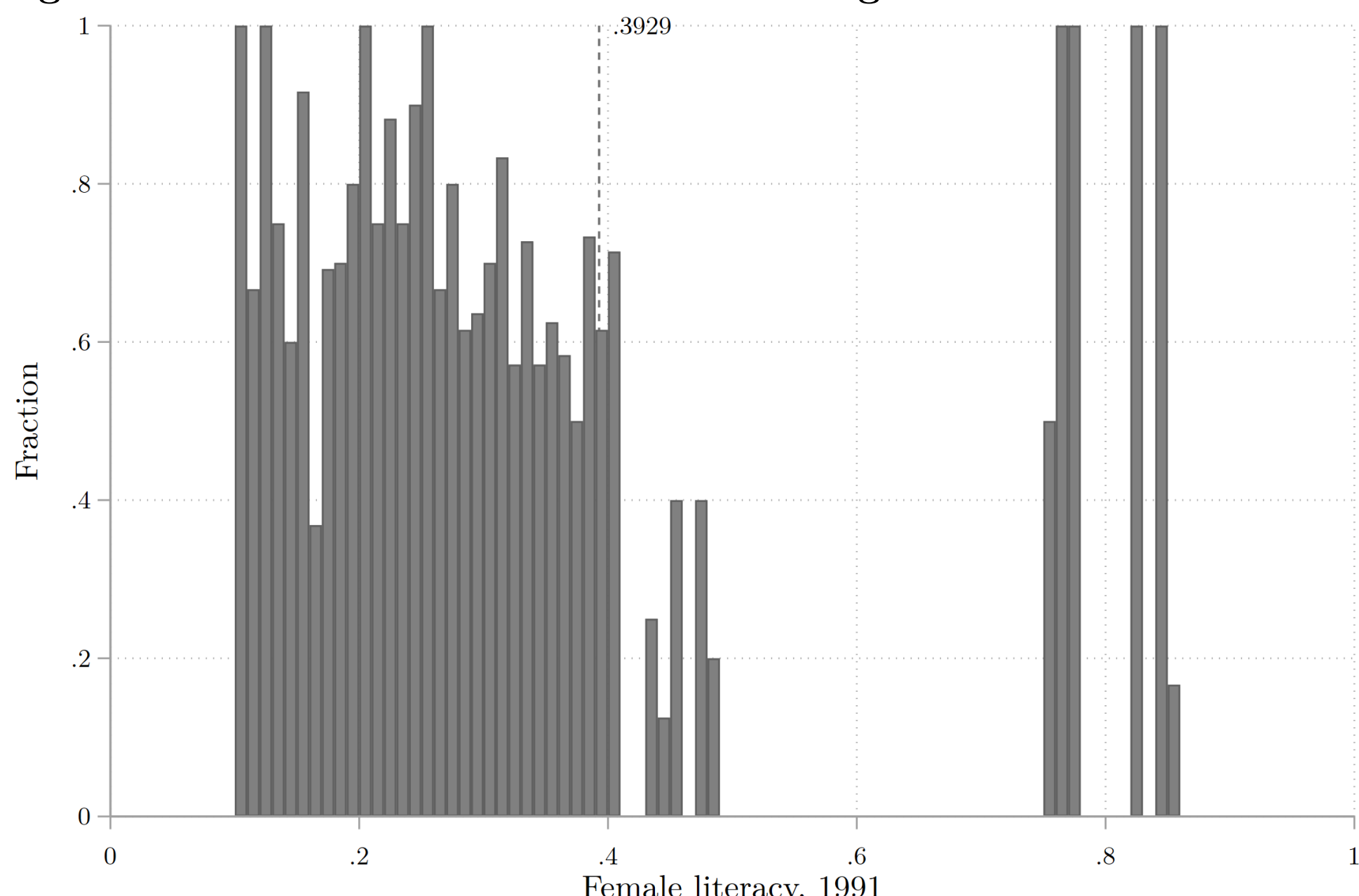

Note: Excludes multi-parent districts using the criterion in footnote 7.

**Table 4. Tests for discontinuities in predetermined variables at various female literacy thresholds**

| Female literacy threshold | Population | Area | Houses | Households | Agricultural worker share |
|---|---|---|---|---|---|
| 0.3929 | 0.93 | 0.57 | 0.69 | 0.71 | 0.40 |
| 0.40 | 0.79 | 0.49 | 0.80 | 0.78 | 1.00 |
| 0.41 | 0.43 | 0.83 | 0.55 | 0.57 | 0.61 |

Notes: Each cell reports a *p* value from a robust RD estimate of the change in a district-level variable observed in 1991 census, at the indicated female literacy threshold. The unit of observation is the 2009 district. The denominator for last the column is number of individuals over age 6. The robust "$\text{HC}_3$" variance estimator is used. Multi-parent districts are excluded.

[16] Khanna (2023b) reports donut RD results at this radius.

**Figure 4. Sensitivity of results from preferred specification to varying the identifying cutoff**

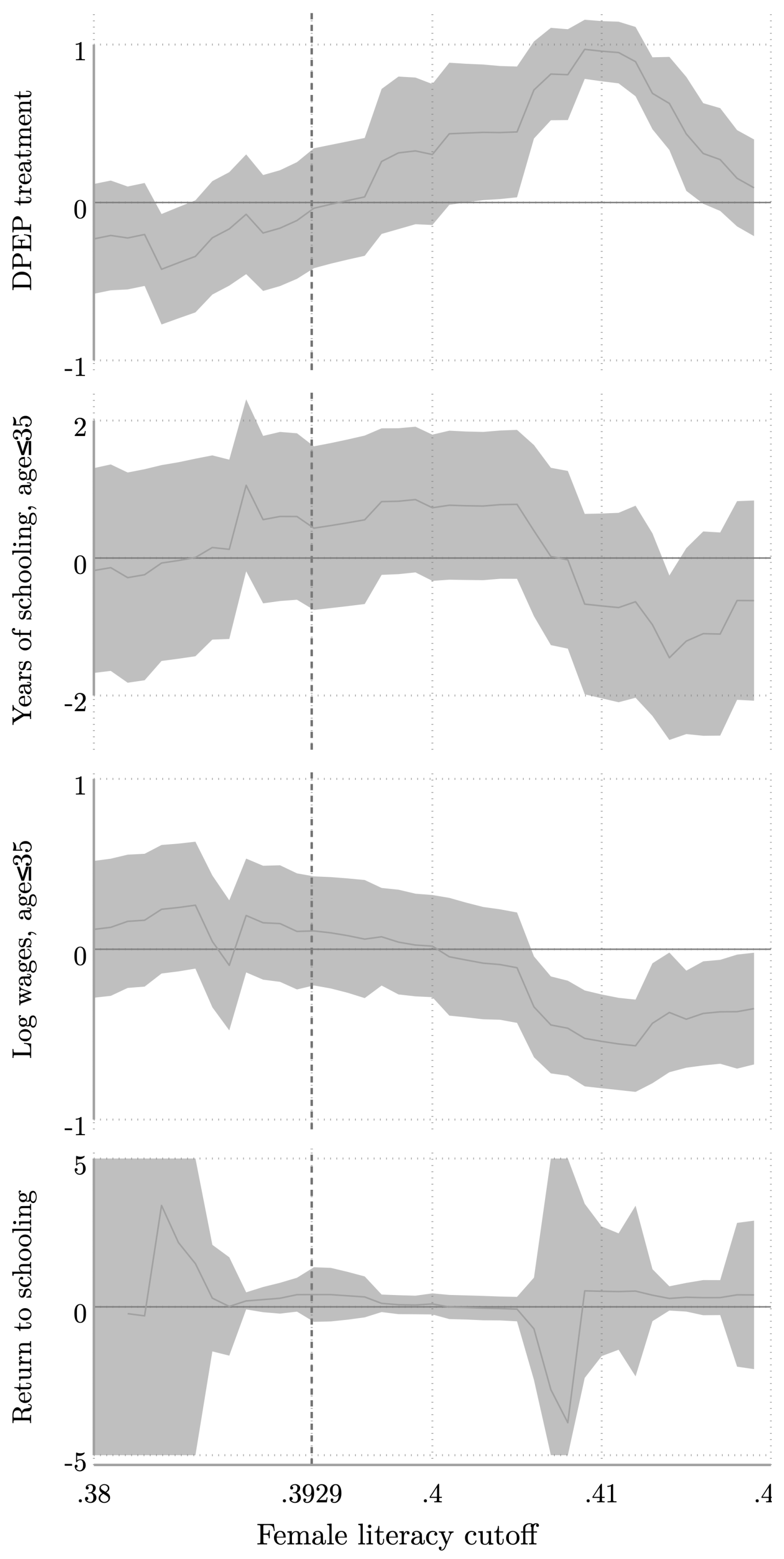

Notes: Each plot shows how the estimated impact on the indicated outcome depends on the cutoff used to define intention to treat. The regressions are based on the specification in the final column of Table 3. Grey regions show 95% confidence intervals. Results in first three plots are sharp RD while those in the last are fuzzy RD. The unit of observation is the district in the first plot and the individual in the rest. For legibility, the vertical range of the bottom plot is clipped to [–5, 5].

**Figure 5. Sensitivity of results from preferred specification to varying the bandwidth**

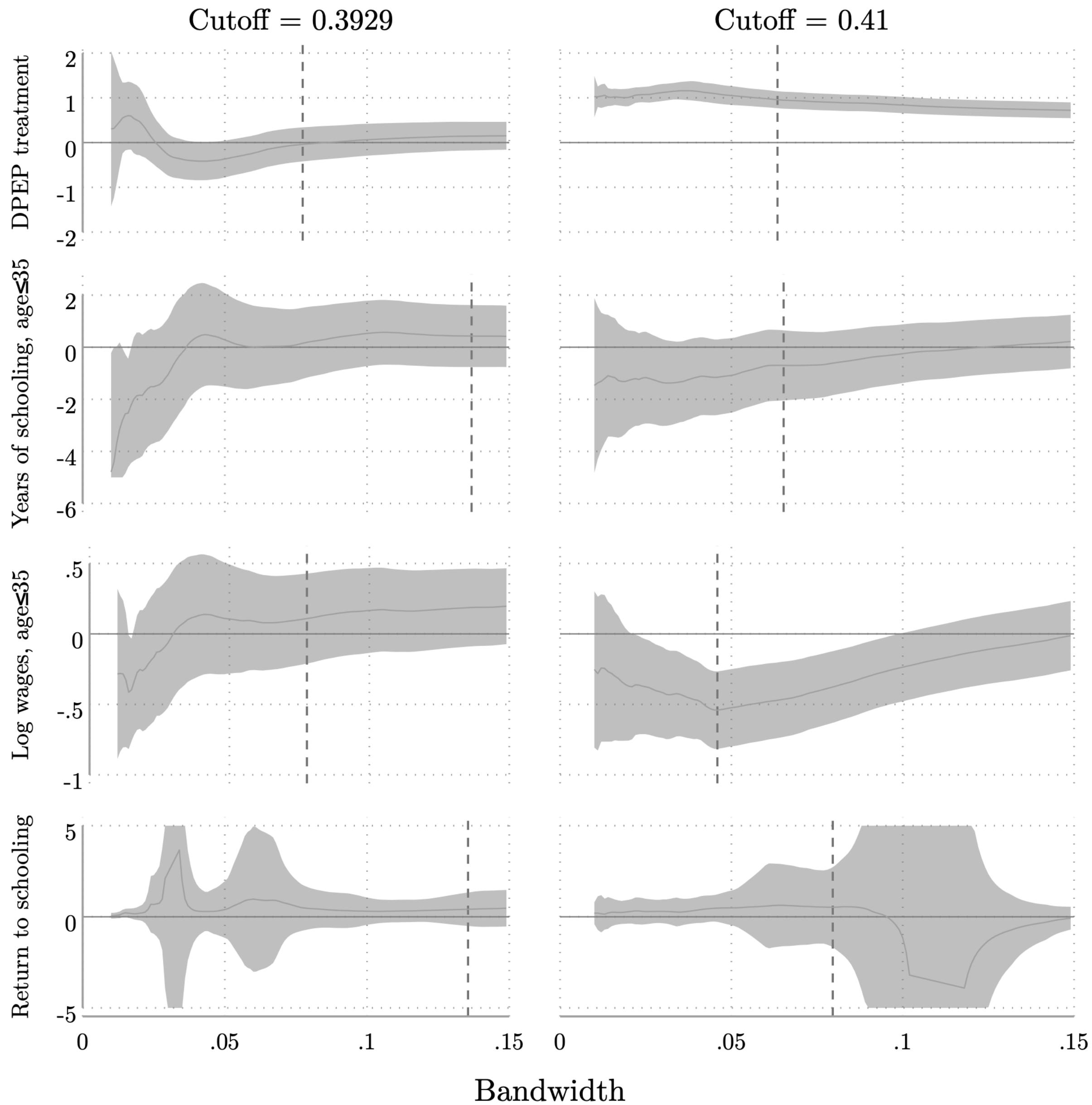

Notes: Plots are constructed much as in the previous figure. The threshold is fixed as indicated and the bandwidth is varied. The vertical lines now show the MSE-optimized bandwidth for each specification, the one the rdrobust program will select by default. Because rdrobust will not separately estimate the bias-correction bandwidth when the estimation bandwidth is manually set, the ratio between the two is chosen to be the same as for when the program is used to automatically choose both.

**Figure 6. Sensitivity of results from preferred specification to varying the bandwidth**

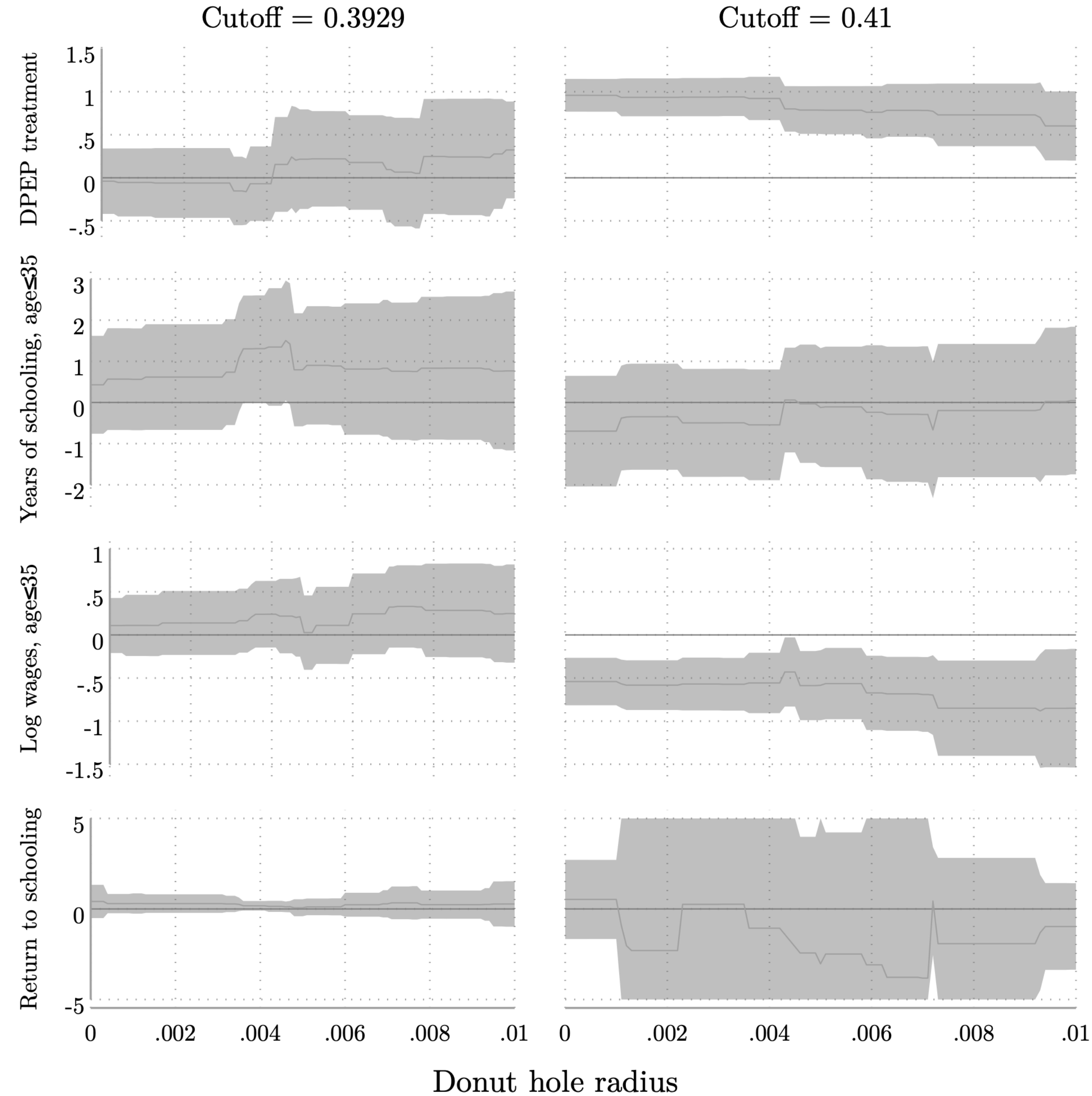

Notes: Plots are constructed much as in the previous figure. After computation of the optimal bandwidth, observations within a given radius of the indicated cutoff are dropped.

## 5 General equilibrium effects

A major contribution of K23 is the way it leverages the large-scale, RD-framed natural experiment to estimate a general equilibrium effects of schooling and then, from there, to estimate elasticities of substitution in production among workers of different age and skill groups. Appendix C describes and comments on the methodology. Here, I check if the GE results are robust to much the same revisions as before.

K23's measurement of the GE effect of DPEP on the skill premium can be constructed as a difference in differences. Using FRD, the K23 code estimates that among young, *unskilled* wage workers near the cutoff (those below age 35 with less than eight years of schooling), those in DPEP districts earned 0.182 log points more. In striking contrast, the corresponding figure for young, *skilled* wage workers is –0.259 log points. In a partial equilibrium perspective, each number is a clean impact estimate: DPEP raised unskilled wages and lowered skilled wages. But the general equilibrium perspective recognizes that the stable unit treatment value assumption is violated. After a large schooling expansion, each skill group's wages are affected by the growth or shrinkage of both groups. To control away some of these effects, K23 benchmarks the impact on skilled wages against the impact on unskilled wages. That is, it subtracts the first number above from the second. This gives a sort of DID estimate of the GE effect on skilled wages: –0.65 log points. To convert that impact on log wages to an impact on log wages per year of schooling, K23 divides by an estimate of the additional schooling attainment among skilled workers, namely, 10 years. That implies that the large-scale economic ripple effects of DPEP cut the skill premium by –0.065 log points per year of schooling. I will label this value $\Delta_D\tilde{\beta}$. It appears in the first columns of Table 1 above and Table 5 below and, subject to a tiny discrepancy, in K23 Table 3.[17]

The K23 methodology for estimating the skill premium in the *absence* of GE effects ($\tilde{\beta}_1$), is more complicated (see appendix C). It too is designed so that certain FRD estimates may be plugged into the formulas. These regressions too estimate impacts of treatment, not ITT. With $\tilde{\beta}_1$ in hand, K23 can compute the relative impact $\Delta_D\tilde{\beta}/\tilde{\beta}_0$ as well as the skill premium in the *presence* of GE effects, $\tilde{\beta}_1 := \tilde{\beta}_0 + \Delta_D\tilde{\beta}$. For inference, K23 bootstraps this entire calculation 1,500 times.

The robustness testing of these computations is reported in Table 5. Results in the first column almost exactly match those in K23's Table 3. The top entry shows that for young adults in non-DPEP districts, the skill premium is equivalent to 0.199 log points per extra year of schooling. In DPEP districts, the premium is lower, 0.134 log points. The difference of 0.065 constitutes a 32% reduction.[18] As in K23, bootstrap $p$ values are displayed in parentheses; they are for a one-sided test that the true sign of the effect is opposite that estimated. As an addition to the K23 reporting, bootstrap percentile-based 95% confidence intervals are also shown, in brackets.[19] The 95% interval around the absolute GE effect of 0.065 is [0.00, 0.19].

[17] K23 Table 3 reports medians of the bootstrap estimates rather than the full-sample estimates.

Column 2 brings in the data revisions other than the inclusion of the missing districts. As before, the outcome and running variables are slightly revised. In addition, the modifications of the DPEP treatment indicator (middle pane of Figure 1) make their first appearance in the numerical analysis, as the treatment variable in the FRD regressions that feed into the GE calculation. Unlike in the RD ITT regressions tested earlier, K23 uses different bandwidths in the various FRD treatment regressions. The bandwidths are recorded in the K23 code and kept fixed across bootstrap replications. How they are chosen is not documented.[20] In the face of this ambiguity, I retain the K23 bandwidths for now, rather than, as in the RD testing, recalculating them for the revised data. The revisions triple the GE effect, to 0.194 log points per year of school. Since the skill premium in non-DPEP districts itself doubles to 0.405, the GE effect relative to that baseline is 48%, up from 32% before.

Here too, adding the four missing districts (column 3) dramatically changes the results. The point estimates shift greatly and are accompanied by far larger confidence intervals. Focusing on $\Delta_D\tilde{\beta}$, whose calculation is simplest, the principal instability is in one of the FRD treatment regressions described at the top of this section, for the impact of DPEP on the pay of skilled, young wage workers. The original estimate of that impact, –0.464 log points per year, is replaced with an implausible +5.93 log points. The problem is that the first stage of that regression is extremely weak. As far as the regression goes, intention to treat hardly affects treatment. The lack of a strong cutoff at 39.29% appears to be combining with the shrinkage of the sample to a particular demographic to weaken the identifying power of the assumed discontinuity.

Subsequent columns cluster the bootstrap by 1991 district (Field and Welsh 2007), switch to optimal bandwidths for each specification and bootstrap replication, apply the CCT bias correction, and incorporate survey weights. The erstwhile penultimate step of introducing CCT's robust standard errors is skipped because it does not affect point estimates, which are all that matter for the bootstrap. None of these revisions alters the finding that the K23 method produces highly unstable results when applied to the full data.

[18] K23 reports 33%. Another source of discrepancy is that K23 takes the ratio of values rounded as displayed.

[19] Like the $p$ values reported in K23, these confidence intervals are constructed directly from the bootstrap distribution, using the 2.5$^{th}$ and 97.5$^{th}$ centiles. "The percentile method is not the last word in bootstrap confidence intervals" (Efron and Hastie 2016). Still, the intuitive method suffices here to indicate the dispersion and skewness of the estimators.

[20] K23's "GE_welfare_bootstrap_part1.do" includes commented-out calls to rdrobust_2014, but when run these produce different bandwidths. The lines do imply that the bandwidths were selected using the CCT method, so that is used here too.

Since the imprecision may owe to the fuzziness of the discontinuity at 39.29%, and since the previous section raised the possibility that the true break was at 41%, I repeat the above analysis using the higher threshold. (See Table 6.) The K23 GE effect estimator is now much more stable, both within each bootstrap (the confidence intervals are narrower) and across data sets and regression specifications. The estimate of $\Delta_D\tilde{\beta}$ starts the testing at 0.033 log points/year and ends at 0.027 without weights and 0.012 with. However, these estimates are all wrong-signed. Now, seemingly, by increasing the supply of skilled workers, DPEP raised their wages. If these results do measure a causal GE effect, it requires another theory for explanation. The result is probably linked to the finding in Figure 4 that when setting the threshold at 41%, the overall impact on wages is negative.

Appendix C proposes revisions to the K23 method of estimating GE effects, as well as the elasticities of substitutions in production between young and old workers, and between the skilled and the unskilled. The results accord with those reported here: imprecise when using the 39.29% cutoff; sharper but wrong-signed when using 41%.

**Table 5. Estimates of general equilibrium effects of DPEP on wages using female literacy cutoff of 39.29%**

| | (1) | (2) | (3) | (4) | (5) | (6) | (7) |
|---|---|---|---|---|---|---|---|
| Skill premium per year of schooling, non-DPEP districts ($\tilde{\beta}_0$) | 0.199 | 0.405 | –2.416 | –2.416 | –0.080 | –0.088 | –7.325 |
| | (0.00) | (0.02) | (0.23) | (0.61) | (0.65) | (0.62) | (0.49) |
| | [0.11, 0.76] | [0.037, 1.80] | [–66.0, 48.8] | [–20.8, 22.2] | [–5.04, 4.82] | [–28.8, 9.49] | [–143, 120] |
| Skill premium per year of schooling, DPEP districts ($\tilde{\beta}_1$) | 0.134 | 0.211 | –1.714 | –1.714 | –0.178 | –0.276 | –9.821 |
| | (0.01) | (0.05) | (0.22) | (0.60) | (0.49) | (0.38) | (0.36) |
| | [0.030, 0.68] | [–0.047, 1.22] | [–44.0, 42.0] | [–20.6, 25.7] | [–6.24, 6.28] | [–41.9, 11.9] | [–423, 74.4] |
| General equilibrium effect ($\Delta_D\tilde{\beta} \coloneqq \tilde{\beta}_1 - \tilde{\beta}_0$) | –0.065 | –0.194 | 0.703 | 0.703 | –0.098 | –0.188 | –2.497 |
| | (0.01) | (0.02) | (0.21) | (0.54) | (0.26) | (0.17) | (0.25) |
| | [–0.19, –0.00] | [–0.62, –0.00] | [–8.76, 9.36] | [–2.79, 3.87] | [–1.70, 2.56] | [–21.0, 2.31] | [–2323, 8.11] |
| General equilibrium effect, % ($\Delta_D\tilde{\beta}/\tilde{\beta}_0$) | –32 | –48 | –29 | –29 | 124 | 213 | 34 |
| | (0.01) | (0.01) | (0.10) | (0.35) | (0.57) | (0.59) | (0.46) |
| | [–77, –3] | [–113, –12] | [–86, 19] | [–492, 679] | [–1083, 1222] | [–3054, 3346] | [–4018, 4343] |
| Revised variables | | ✓ | ✓ | ✓ | ✓ | ✓ | ✓ |
| Four missing districts | | | ✓ | ✓ | ✓ | ✓ | ✓ |
| Clustered | | | | ✓ | ✓ | ✓ | ✓ |
| Regression-specific BW | | | | | ✓ | ✓ | ✓ |
| Bias-corrected | | | | | | ✓ | ✓ |
| Weights | | | | | | | ✓ |

*Notes*: This table is structured much like Table 3, starting with the original estimates in column 1 and cumulating specification changes. However, since moving to CCT robust standard errors, as in column 11 in Table 3, does not change point estimates, it does not affect the bootstrap distribution either, so the step is skipped. Bootstrap $p$ values are in parentheses, for the one-sided test that a parameter has the sign opposite that estimated. Percentile-based bootstrap 95% confidence intervals are in brackets.

**Table 6. Estimates of general equilibrium effects of DPEP on wages using female literacy cutoff of 41%**

| | (1) | (2) | (3) | (4) | (5) | (6) | (7) |
|---|---|---|---|---|---|---|---|
| Skill premium per year of schooling, non-DPEP districts ($\tilde{\beta}_0$) | 0.360 | 0.428 | 0.252 | 0.252 | 0.100 | 0.113 | 0.169 |
| | (0.09) | (0.05) | (0.00) | (0.16) | (0.30) | (0.24) | (0.34) |
| | [–2.89, 3.85] | [–2.36, 3.20] | [0.13, 0.72] | [–1.94, 2.57] | [–0.65, 0.66] | [–0.62, 0.57] | [–1.80, 1.45] |
| Skill premium per year of schooling, DPEP districts ($\tilde{\beta}_1$) | 0.393 | 0.455 | 0.287 | 0.287 | 0.138 | 0.140 | 0.181 |
| | (0.09) | (0.05) | (0.00) | (0.12) | (0.23) | (0.19) | (0.29) |
| | [–2.85, 3.89] | [–2.32, 3.23] | [0.17, 0.77] | [–1.88, 2.58] | [–0.68, 0.72] | [–0.61, 0.61] | [–1.81, 1.47] |
| General equilibrium effect ($\Delta_D\tilde{\beta} \coloneqq \tilde{\beta}_1 - \tilde{\beta}_0$) | 0.033 | 0.027 | 0.035 | 0.035 | 0.038 | 0.027 | 0.012 |
| | (0.03) | (0.06) | (0.01) | (0.07) | (0.10) | (0.16) | (0.22) |
| | [–0.002, 0.09] | [–0.008, 0.061] | [0.005, 0.066] | [–0.001, 0.13] | [–0.023, 0.087] | [–0.029, 0.072] | [–0.054, 0.096] |
| General equilibrium effect, % ($\Delta_D\tilde{\beta}/\tilde{\beta}_0$) | 9 | 6 | 14 | 14 | 38 | 24 | 7 |
| | (0.11) | (0.10) | (0.01) | (0.20) | (0.34) | (0.32) | (0.41) |
| | [–4, 33] | [–3, 25] | [1, 44] | [–179, 233] | [–464, 495] | [–448, 359] | [–345, 304] |
| Revised variables | | ✓ | ✓ | ✓ | ✓ | ✓ | ✓ |
| Four missing districts | | | ✓ | ✓ | ✓ | ✓ | ✓ |
| Clustered | | | | ✓ | ✓ | ✓ | ✓ |
| Regression-specific BW | | | | | ✓ | ✓ | ✓ |
| Bias-corrected | | | | | | ✓ | ✓ |
| Weights | | | | | | | ✓ |

*Notes*: See notes to previous table. The methodology used here differs only in moving the assumed discontinuity from 39.29% to 41%

## 6 Conclusion

K23 highlights how general equilibrium dynamics mediate the impacts of government initiatives. It proposes a method for distinguishing partial- and general-equilibrium effects in the context of RD-based program evaluation. And it applies these ideas to an externally financed education program in India in the 1990s. This comment emphasizes the empirics. The estimates of impacts, GE effects, and elasticities of substitution are not robust to changes in data and method that ought to improve inference. The inclusion of four missing districts and the clustering of standard errors matter most. The revised estimates are hard to distinguish from zero. These findings are robust to varying the identifying threshold, the bandwidth, and the radius of an exclusionary donut hole. One deep source of the uncertainty is the lack of a strong discontinuity in treatment right at the ascribed cutoff of 39.29%. GE estimates using the cutoff of 41% are more stable, but wrong-signed, thus hard to interpret. Since the Indian program only increased school spending 17.5–20% for 5–7 years in affected districts, the difficulty of detecting impacts should perhaps not surprise.

## References

Abadie, Alberto, Susan Athey, Guido W. Imbens, and Jeffrey M. Wooldridge. 2022. "When Should You Adjust Standard Errors for Clustering?" *Quarterly Journal of Economics* 138 (1): 1–35. https://doi.org/10.1093/qje/qjac038.

Almond, Douglas, and Joseph J. Doyle. 2011. "After Midnight: A Regression Discontinuity Design in Length of Postpartum Hospital Stays." *American Economic Journal: Economic Policy* 3 (3): 1–34. https://doi.org/10.1257/pol.3.3.1.

Azam, Mehtabul, and Chan Hang Saing. 2017. "Assessing the Impact of District Primary Education Program in India." *Review of Development Economics* 21 (4): 1113–31. https://doi.org/10.1111/rode.12281.

Barreca, Alan I., Jason M. Lindo, and Glen R. Waddell. 2016. "Heaping-Induced Bias in Regression-Discontinuity Designs." *Economic Inquiry* 54 (1): 286–93. https://doi.org/10.1111/ecin.12225.

Bartalotti, Otávio, and Quentin Brummet. 2017. "Regression discontinuity designs with clustered data." In *Advances in Econometrics: Vol. 38—Regression Discontinuity Designs: Theory and Applications*, eds. M. D. Cattaneo and J. C. Escanciano, 383–420. Emerald. https://doi.org/10.1108/S0731-905320170000038017.

Bertrand, Marianne, Esther Duflo, and Sendhil Mullainathan. 2004. "How Much Should We Trust Differences-In-Differences Estimates?" *Quarterly Journal of Economics* 119 (1): 249–75. https://doi.org/10.1162/003355304772839588.

Calonico, Sebastian, Matias D. Cattaneo, and Max H. Farrell. 2018. "Coverage Error Optimal Confidence Intervals for Local Polynomial Regression." *arXiv [econ.EM]*. http://arxiv.org/abs/1808.01398.

Calonico, Sebastian, Matias D. Cattaneo, Max H. Farrell, and Rocío Titiunik. 2017. "rdrobust: Software for Regression-Discontinuity Designs." *Stata Journal* 17 (2): 372–404. https://doi.org/10.1177/1536867X1701700208.

Calonico, Sebastian, Matias D. Cattaneo, Max H. Farrell, and Rocío Titiunik. 2019. "Regression Discontinuity Designs Using Covariates." *Review of Economics and Statistics* 101 (3): 442–51. https://doi.org/10.1162/rest_a_00760.

Calonico, Sebastian, Matias D. Cattaneo, and Rocio Titiunik. 2014a. "Robust Nonparametric Confidence Intervals for Regression-Discontinuity Designs." *Econometrica* 82 (6): 2295–2326. https://doi.org/10.3982/ecta11757.

Calonico, Sebastian, Matias D. Cattaneo, and Rocío Titiunik. 2014b. "Robust Data-Driven Inference in the Regression-Discontinuity Design." *Stata Journal* 14 (4): 909–46.

https://doi.org/10.1177/1536867X1401400413.

Calonico, Sebastian, Matias D. Cattaneo, and Rocío Titiunik. 2015. "Optimal Data-Driven Regression Discontinuity Plots." *Journal of the American Statistical Association* 110 (512): 1753–69. https://doi.org/10.1080/01621459.2015.1017578.

Cattaneo, Matias D., Nicolás Idrobo, and Rocío Titiunik. 2019. "A Practical Introduction to Regression Discontinuity Designs: Foundations." In *Elements in Quantitative and Computational Methods for the Social Sciences*. Cambridge University Press. https://doi.org/10.1017/9781108684606.

Deaton, Angus. 1997. *The Analysis of Household Surveys: A Microeconometric Approach to Development Policy*. World Bank Publications. https://hdl.handle.net/10986/30394.

Department of Education, Ministry of Human Resource Development, India. 1995. *District Primary Education Programme: Guidelines*. https://scribd.com/document/410912918/DPEP-guidelines-pdf.

Duflo, Esther. 2001. "Schooling and Labor Market Consequences of School Construction in Indonesia: Evidence from an Unusual Policy Experiment." *American Economic Review* 91 (4): 795–813. https://doi.org/10.1257/aer.91.4.795.

Duflo, Esther, Pascaline Dupas, and Michael Kremer. 2021. "The Impact of Free Secondary Education: Experimental Evidence from Ghana." Working Paper. National Bureau of Economic Research. https://doi.org/10.3386/w28937.

Efron, Bradley, and Trevor Hastie. 2016. *Computer Age Statistical Inference: Algorithms, Evidence, and Data Science*. Cambridge University Press. https://cambridge.org/9781107149892.

Field, C. A., and Alan H. Welsh. 2007. "Bootstrapping Clustered Data." *Journal of the Royal Statistical Society. Series B, Statistical Methodology* 69 (3): 369–90. https://doi.org/10.1111/j.1467-9868.2007.00593.x.

Gelman, Andrew, and Guido Imbens. 2019. "Why High-Order Polynomials Should Not Be Used in Regression Discontinuity Designs." *Journal of Business & Economic Statistics* 37 (3): 447–56. https://doi.org/10.1080/07350015.2017.1366909.

Hausman, Jerry A., and David A. Wise. 1981. "Stratification on an Endogenous Variable and Estimation: The Gary Income Maintenance Experiment." In Charles F. Manski and Daniel L. McFadden, eds. *Structural Analysis of Discrete Data with Econometric Applications*. MIT Press.

Imbens, Guido W., and Joshua D. Angrist. 1994. "Identification and Estimation of Local Average Treatment Effects." *Econometrica* 62 (2): 467–75. https://doi.org/10.2307/2951620.

Imbens, Guido, and Karthik Kalyanaraman. 2012. "Optimal Bandwidth Choice for the Regression Discontinuity Estimator." *Review of Economic Studies* 79 (3): 933–59.

https://doi.org/10.1093/restud/rdr043.

Jalan, Jyotsna, and Elena Glinskaya. 2013. "Small Bang for Big Bucks? An Evaluation of a Primary School Intervention in India." Centre for Training and Research in Public Finance and Policy. https://doi.org/10.13140/RG.2.1.2978.7040.

Khanna, Gaurav. 2023a. "Large-Scale Education Reform in General Equilibrium: Regression Discontinuity Evidence from India." *Journal of Political Economy* 131 (2): 541–91 https://doi.org/10.1086/721619.

Khanna, Gaurav. 2023b. "Response to 'Comment on Khanna (2023)'." Discussion Paper 71. Institute for Replication. https://www.rwi-essen.de/i4r-discussion-papers-series-1/response-to-comment-on-khanna-2023-23092102.

Kinal, Terrence W. 1980. "The Existence of Moments of *k*-Class Estimators." *Econometrica* 48 (1): 241–49. https://doi.org/10.2307/1912027.

Kolesár, Michal, and Christoph Rothe. 2018. "Inference in Regression Discontinuity Designs with a Discrete Running Variable." *American Economic Review* 108 (8): 2277–2304. https://doi.org/10.1257/aer.20160945.

Kuipers, Nicholas, Gareth Nellis, and Drew Stommes. 2025. "Forging Social Cohesion Through Mass Education: Evidence from a Nationwide Policy Reform in India." https://researchgate.net/publication/390272160_Forging_Social_Cohesion_Through_Mass_Education_Evidence_from_a_Nationwide_Policy_Reform_in_India.

Kumar, Hemanshu, and Rohini Somanathan. 2016. "Creating Long Panels Using Census Data: 1961–2001." cdedse.org/pdf/work248.pdf.

MacKinnon, James G., and Halbert White. 1985. "Some Heteroskedasticity-Consistent Covariance Matrix Estimators with Improved Finite Sample Properties." *Journal of Econometrics* 29 (3): 305–25. https://doi.org/10.1016/0304-4076(85)90158-7.

Minnesota Population Center. 2020. Integrated Public Use Microdata Series, International: Version 7.3 [dataset]. https://doi.org/10.18128/D020.V7.3

Over, Mead. 2022. "grc1leg2. Add a common legend to combined graphs." http://digital.cgdev.org/doc/stata/MO/Misc.

Pandey, Raghaw Sharan. 2000. "Going to Scale with Education Reform: India's District Primary Education Program, 1995–99." Country Studies. Education Reform and Management Publication Series I (4). World Bank. http://web.worldbank.org/archive/website00238I/WEB/PDF/INDIA.PDF.

Pisati, Maurizio. 2004. "Simple Thematic Mapping." *Stata Journal* 4 (4): 361–78.

https://doi.org/10.1177/1536867X0400400401.

Planning Commission, Government of India. 1994. *Annual Plan 1994–95*. https://planningcommission.gov.in/docs/reports/publications/anpl_9495e.pdf. Accessed July 1, 2022.

Potter, Frank, and Yuhong Zheng. 2015. "Methods and Issues in Trimming Extreme Weights in Sample Surveys." Proceedings of the Joint Statistical Meetings Survey Research Methods Section. American Statistical Association. http://www.asasrms.org/Proceedings/y2015/files/234115.pdf.

Roodman, David. [2026]. "Schooling and Labor Market Consequences of School Construction in Indonesia: Comment." [*Journal of Comments and Replications in Economics*]

Simmons, Joseph P., Leif D. Nelson, and Uri Simonsohn. 2011. "False-Positive Psychology: Undisclosed Flexibility in Data Collection and Analysis Allows Presenting Anything as Significant." *Psychological Science* 22 (11): 1359–66. https://doi.org/10.1177/0956797611417632.

Solon, Gary, Steven J. Haider, and Jeffrey M. Wooldridge. 2015. "What Are We Weighting For?" *Journal of Human Resources* 50 (2): 301–16. https://doi.org/10.3368/jhr.50.2.301.

Sunder, Naveen. 2018. "Parents' Schooling & Intergenerational Human Capital: Evidence from India." https://www.aeaweb.org/conference/2019/preliminary/paper/9NiR5sBe.

Van de Kerckhove, Wendy, Leyla Mohadjer, and Thomas Krenzke 2014. "A Weight Trimming Approach to Achieve a Comparable Increase to Bias across Countries in the Programme for the International Assessment of Adult Competencies." Proceedings of the Joint Statistical Meetings Survey Research Methods Section. American Statistical Association. http://www.asasrms.org/Proceedings/y2014/files/311170_87007.pdf.

Yon, George Vega. 2012 "Introducing PARALLEL: Stata Module for Parallel Computing." http://fmwww.bc.edu/repec/bocode/p/parallel.pdf.

# Appendices

## A Disagreements in female literacy assignment

Table A–1 below documents where the K23 and new data sets disagree on the values of female literacy (FL) assigned to India's districts as they were defined in 2009. The literacy values come from the 1991 census. Because some "multi-parent" 2009 districts were formed out of mergers of parts of 1991 districts, it is ambiguous what FL values to assign them. In the new data set, multi-parent districts are only assigned a value when the population contributions of all but one parent are *de minimus* or when the parents have nearly the same FL. The formal criterion is that the standard deviation of parental FL, weighting by their population contributions to the child, must be less than 0.01. Population contributions for 1991–2001 district changes are taken from Kumar and Somanathan (2016, Tables 7, 8, 9d); for 2001–09 they computed from official reports from the 2011 census.

The table is structured to exhibit the data behind these calculations, in cases of disagreement.

The first row shows the sole case of an apparent data entry error in the K23 data: East Nimar's FL of 31.53% is used for West Nimar instead of the correct value of 23.23%.

The next four rows show the districts that are fully missing in K23. The periods in the "K23 FL" column indicate missing values.

Next are listed 8 single-parent districts that have different names from their parents, and yet which, being single-parent, can be unambiguously assigned 1991 FL values—and are not in K23.

Then come the three districts that are assigned FL values in the new data but not in K23, followed by 13 in the opposite situation. Each of these districts is represented with multiple rows in order to document their parentage. In the first example, Nawanshahr district in Punjab, both parents made substantial population contributions (71.1% and 28.9%) but have nearly identical 1991 FL rates, 61.33% and 61.48%. The standard deviation of parental FL is only 0.069%, which satisfies the *de minimus* criterion. In the new data, the child district is assigned a (population-weighted) FL of 61.37%.

Lower down is the case of Sonipat, Haryana. Though born of a nearly identical population split—72.2%/27.8%—it is given opposite treatment in both data sets. Its standard deviation of parental FL lands just above the 1% threshold so it is assigned no FL value in the new data. But K23 assigns it the FL value of the larger parent. It is unclear what the criterion was in K23 for assigning FL values to multiparent districts such as Sonipat and Nawanshahr.

## Table A–1. Disagreements in female literacy (FL) assignments between K23 and this reanalysis

| 2009 district code | 2009 State | 2009 District | K23 FL | New FL | Standard deviation of parental FL | 1991 State | 1991 District | Share of new from old | 1991 FL |
|---|---|---|---|---|---|---|---|---|---|
| *Erroneous value assigned in K23 (from East Nimar)* | | | | | | | | | |
| 2327 | Madhya Pradesh | W. Nimar | 31.53% | 23.23% | 0.000% | Madhya Pradesh | West Nimar | 100.0% | 23.23% |
| *Fully absent from K23 data set* | | | | | | | | | |
| 2719 | Maharashtra | Aurangabad | . | 39.64% | 0.000% | Maharashtra | Aurangabad | 100.0% | 39.64% |
| 1402 | Manipur | Tamenglong | . | 39.68% | 0.000% | Manipur | Tamenglong | 100.0% | 39.68% |
| 3318 | Tamil Nadu | Cuddalore | . | 39.70% | 0.000% | Tamil Nadu | South Arcot | 100.0% | 39.70% |
| 2728 | Maharashtra | Latur | . | 39.74% | 0.000% | Maharashtra | Latur | 100.0% | 39.74% |
| *Single-parent split districts present in K23 data set but assigned no FL value* | | | | | | | | | |
| 1826 | Assam | Guwahati | . | 55.01% | 0.000% | Assam | Kamrup | 100.0% | 55.01% |
| 1827 | Assam | Udalguri | . | 32.53% | 0.000% | Assam | Darrang | 100.0% | 32.53% |
| 610 | Haryana | Fatehabad | . | 32.12% | 0.000% | Haryana | Hisar | 100.0% | 32.12% |
| 2003 | Jharkhand | Chatra | . | 21.24% | 0.000% | Bihar | Hazaribag | 100.0% | 21.24% |
| 2005 | Jharkhand | Kodarma | . | 21.24% | 0.000% | Bihar | Hazaribag | 100.0% | 21.24% |
| 2022 | Jharkhand | Saraikela Khareswan | . | 22.44% | 0.000% | Bihar | Pashchimi Singhbhum | 100.0% | 22.44% |
| 2123 | Orissa | Sonapur | . | 21.88% | 0.000% | Orissa | Balangir | 100.0% | 21.88% |
| 3331 | Tamil Nadu | Krishnagiri | . | 34.23% | 0.000% | Tamil Nadu | Dharmapuri | 100.0% | 34.23% |
| *Multi-parent districts with FL value in K23, but no value in new data set (one 2009 district per row group)* | | | | | | | | | |
| 306 | Punjab | Nawanshahr | . | 61.37% | 0.069% | Punjab | Jalandhar | 71.1% | 61.33% |
| 306 | Punjab | Nawanshahr | . | 61.37% | 0.069% | Punjab | Hoshiarpur | 28.9% | 61.48% |
| 2208 | Chhattisgarh | Kawardha | . | 27.65% | 0.267% | Madhya Pradesh | Rajnandgaon | 68.3% | 27.83% |
| 2208 | Chhattisgarh | Kawardha | . | 27.65% | 0.267% | Madhya Pradesh | Bilaspur | 31.7% | 27.26% |
| 2420 | Gujarat | Narmada | . | 49.97% | 0.738% | Gujarat | Bharuch | 88.5% | 49.71% |
| 2420 | Gujarat | Narmada | . | 49.97% | 0.738% | Gujarat | Vadodara | 11.5% | 52.02% |
| *Multi-parent districts with FL value in K23, but given no value in new data set (one 2009 district per row group)* | | | | | | | | | |
| 609 | Haryana | Jind | 30.12% | . | 1.093% | Haryana | Jind | 96.4% | 30.12% |
| 609 | Haryana | Jind | 30.12% | . | 1.093% | Haryana | Kaithal | 3.2% | 28.37% |
| 609 | Haryana | Jind | 30.12% | . | 1.093% | Haryana | Rohtak | 0.4% | 45.74% |
| 608 | Haryana | Sonipat | 48.27% | . | 1.132% | Haryana | Sonipat | 72.2% | 48.27% |
| 608 | Haryana | Sonipat | 48.27% | . | 1.132% | Haryana | Rohtak | 27.8% | 45.74% |
| 612 | Haryana | Hisar | 32.12% | . | 1.150% | Haryana | Hisar | 99.1% | 32.12% |

| 2009 district code | 2009 State | 2009 District | K23 FL | New FL | Standard deviation of parental FL | 1991 State | 1991 District | Share of new from old | 1991 FL |
|---|---|---|---|---|---|---|---|---|---|
| 612 | Haryana | Hisar | 32.12% | . | 1.150% | Haryana | Rohtak | 0.7% | 45.74% |
| 612 | Haryana | Hisar | 32.12% | . | 1.150% | Haryana | Bhiwani | 0.2% | 35.10% |
| 605 | Haryana | Kaithal | 28.37% | . | 1.323% | Haryana | Kaithal | 98.2% | 28.37% |
| 605 | Haryana | Kaithal | 28.37% | . | 1.323% | Haryana | Jind | 1.3% | 30.12% |
| 605 | Haryana | Kaithal | 28.37% | . | 1.323% | Haryana | Kurukshetra | 0.5% | 46.94% |
| 613 | Haryana | Bhiwani | 35.10% | . | 1.567% | Haryana | Bhiwani | 97.8% | 35.10% |
| 613 | Haryana | Bhiwani | 35.10% | . | 1.567% | Haryana | Rohtak | 2.2% | 45.74% |
| 2413 | Gujarat | Amreli | 48.77% | . | 1.666% | Gujarat | Amreli | 83.0% | 48.77% |
| 2413 | Gujarat | Amreli | 48.77% | . | 1.666% | Gujarat | Bhavnagar | 17.0% | 44.33% |
| 956 | Uttar Pradesh | S. Kabir Nagar | 17.82% | . | 1.683% | Uttar Pradesh | Basti | 91.3% | 17.82% |
| 956 | Uttar Pradesh | S. Kabir Nagar | 17.82% | . | 1.683% | Uttar Pradesh | Siddharth Nagar | 8.7% | 11.84% |
| 606 | Haryana | Karnal | 43.54% | . | 1.972% | Haryana | Karnal | 83.7% | 43.54% |
| 606 | Haryana | Karnal | 43.54% | . | 1.972% | Haryana | Panipat | 14.7% | 41.17% |
| 606 | Haryana | Karnal | 43.54% | . | 1.972% | Haryana | Jind | 0.8% | 30.12% |
| 606 | Haryana | Karnal | 43.54% | . | 1.972% | Haryana | Kaithal | 0.8% | 28.37% |
| 309 | Punjab | Ludhiana | 61.23% | . | 2.282% | Punjab | Ludhiana | 98.8% | 61.23% |
| 309 | Punjab | Ludhiana | 61.23% | . | 2.282% | Punjab | Sangrur | 1.0% | 37.86% |
| 309 | Punjab | Ludhiana | 61.23% | . | 2.282% | Punjab | Hoshiarpur | 0.1% | 61.48% |
| 309 | Punjab | Ludhiana | 61.23% | . | 2.282% | Punjab | Patiala | 0.1% | 50.33% |
| 309 | Punjab | Ludhiana | 61.23% | . | 2.282% | Punjab | Jalandhar | 0.1% | 61.33% |
| 604 | Haryana | Kurukshetra | 46.94% | . | 2.717% | Haryana | Kurukshetra | 95.3% | 46.94% |
| 604 | Haryana | Kurukshetra | 46.94% | . | 2.717% | Haryana | Yamunanagar | 2.3% | 50.07% |
| 604 | Haryana | Kurukshetra | 46.94% | . | 2.717% | Haryana | Kaithal | 2.0% | 28.37% |
| 604 | Haryana | Kurukshetra | 46.94% | . | 2.717% | Haryana | Karnal | 0.3% | 43.54% |
| 947 | Uttar Pradesh | Faizabad | 22.97% | . | 2.930% | Uttar Pradesh | Faizabad | 81.6% | 22.97% |
| 947 | Uttar Pradesh | Faizabad | 22.97% | . | 2.930% | Uttar Pradesh | Barabanki | 18.4% | 15.41% |
| 2305 | Madhya Pradesh | Datia | 23.69% | . | 7.600% | Madhya Pradesh | Datia | 76.9% | 23.69% |
| 2305 | Madhya Pradesh | Datia | 23.69% | . | 7.600% | Madhya Pradesh | Gwalior | 23.1% | 41.72% |
| 934 | Uttar Pradesh | Kanpur Nagar | 58.82% | . | 10.003% | Uttar Pradesh | Kanpur Nagar | 74.3% | 58.82% |
| 934 | Uttar Pradesh | Kanpur Nagar | 58.82% | . | 10.003% | Uttar Pradesh | Kanpur Dehat | 25.7% | 35.92% |

Notes: "K23 FL" and "New FL" are the 1991 female district-level female literacy rates assigned to 2009 districts in K23 and this reanalysis, with "." indicating no assignment. "Standard deviation of parental FL" is the standard deviation of the female literacy rates of 1991 parent districts, weighting by their population contributions to child districts, which are in the "Share of old from new" column.

## B Reanalysis of RD results using female literacy cutoff of 41%

**Table B–1. Replication and revision of Khanna (2023a) RD regressions**

| | (1) | (2) | (3) | (4) | (5) | (6) | (7) | (8) | (9) | (10) | (11) | (12) |
|---|---|---|---|---|---|---|---|---|---|---|---|---|
| *Impact of intention to treat on schooling: young (sharp RD)* | | | | | | | | | | | | |
| Estimate | –0.702$^{**}$ | –0.250 | –0.701$^{**}$ | –0.360 | –0.211 | 0.170 | –0.211 | 0.170 | –0.080 | –0.240 | –0.240 | –0.549 |
| | (0.280) | (0.209) | (0.273) | (0.221) | (0.244) | (0.189) | (0.538) | (0.419) | (0.477) | (0.477) | (0.554) | (0.661) |
| Bandwidth | 0.054 | 0.100 | 0.055 | 0.089 | 0.067 | 0.114 | 0.067 | 0.114 | 0.086 | 0.086 | 0.086 | 0.073 |
| Observations | 5,849 | 10,524 | 5,934 | 8,929 | 7,135 | 12,108 | 7,135 | 12,108 | 8,737 | 8,737 | 8,737 | 7,646 |
| *Impact of intention to treat on log wages: young (sharp RD)* | | | | | | | | | | | | |
| Estimate | –0.243$^{***}$ | –0.097$^{***}$ | –0.273$^{***}$ | –0.155$^{***}$ | –0.238$^{***}$ | –0.035 | –0.238$^{**}$ | –0.035 | –0.241$^{**}$ | –0.314$^{**}$ | –0.314$^{**}$ | –0.465$^{***}$ |
| | (0.043) | (0.032) | (0.044) | (0.035) | (0.039) | (0.030) | (0.110) | (0.092) | (0.123) | (0.123) | (0.137) | (0.141) |
| Bandwidth | 0.054 | 0.100 | 0.055 | 0.089 | 0.067 | 0.114 | 0.067 | 0.114 | 0.054 | 0.054 | 0.054 | 0.048 |
| Observations | 5,849 | 10,524 | 5,934 | 8,929 | 7,135 | 12,108 | 7,135 | 12,108 | 6,198 | 6,198 | 6,198 | 5,625 |
| *Impact of schooling on log wages: young (fuzzy RD)* | | | | | | | | | | | | |
| Estimate | 0.346$^{***}$ | 0.390 | 0.390$^{***}$ | 0.430$^{*}$ | 1.126 | –0.205 | 1.126 | –0.205 | –1.774 | –9.309 | –9.309 | 0.648 |
| | (0.118) | (0.283) | (0.132) | (0.232) | (1.223) | (0.350) | (2.781) | (0.852) | (14.188) | (14.188) | (16.854) | (1.453) |
| Bandwidth | 0.054 | 0.100 | 0.055 | 0.089 | 0.067 | 0.114 | 0.067 | 0.114 | 0.094 | 0.094 | 0.094 | 0.078 |
| Observations | 5,849 | 10,524 | 5,934 | 8,929 | 7,135 | 12,108 | 7,135 | 12,108 | 10,320 | 10,320 | 10,320 | 8,185 |
| *Impact of intention to treat on schooling: old (sharp RD)* | | | | | | | | | | | | |
| Estimate | 0.007 | –0.031 | 0.035 | –0.038 | 0.053 | 0.273 | 0.053 | 0.273 | 0.199 | 0.145 | 0.145 | 0.206 |
| | (0.295) | (0.221) | (0.291) | (0.234) | (0.258) | (0.200) | (1.018) | (0.712) | (0.768) | (0.768) | (0.930) | (0.791) |
| Bandwidth | 0.054 | 0.100 | 0.055 | 0.089 | 0.067 | 0.114 | 0.067 | 0.114 | 0.099 | 0.099 | 0.099 | 0.107 |
| Observations | 6,701 | 12,055 | 6,798 | 10,292 | 8,080 | 13,912 | 8,080 | 13,912 | 12,425 | 12,425 | 12,425 | 13,471 |
| *Impact of intention to treat on log wages: old (sharp RD)* | | | | | | | | | | | | |
| Estimate | –0.050 | –0.024 | –0.052 | –0.028 | –0.092$^{**}$ | 0.036 | –0.092 | 0.036 | –0.004 | –0.056 | –0.056 | –0.140 |
| | (0.053) | (0.038) | (0.052) | (0.041) | (0.045) | (0.035) | (0.203) | (0.150) | (0.161) | (0.161) | (0.191) | (0.186) |
| Bandwidth | 0.054 | 0.100 | 0.055 | 0.089 | 0.067 | 0.114 | 0.067 | 0.114 | 0.099 | 0.099 | 0.099 | 0.078 |
| Observations | 6,699 | 12,052 | 6,798 | 10,292 | 8,080 | 13,912 | 8,080 | 13,912 | 12,425 | 12,425 | 12,425 | 9,262 |
| *Impact of schooling on log wages: old (fuzzy RD)* | | | | | | | | | | | | |
| Estimate | –12.1 | 0.675 | –1.480 | 0.735 | –1.749 | 0.134 | –1.749 | 0.134 | 0.010 | –0.244 | –0.244 | –0.253 |
| | (875) | (3.530) | (13.239) | (3.861) | (9.196) | (0.091) | (37.424) | (0.281) | (0.726) | (0.726) | (0.868) | (0.665) |
| Bandwidth | 0.054 | 0.100 | 0.055 | 0.089 | 0.067 | 0.114 | 0.067 | 0.114 | 0.101 | 0.101 | 0.101 | 0.101 |
| Observations | 6,699 | 12,052 | 6,798 | 10,292 | 8,080 | 13,912 | 8,080 | 13,912 | 12,643 | 12,643 | 12,643 | 12,662 |
| Bandwidth method | CCT | IK | CCT | IK | CCT | IK | CCT | IK | CCT | CCT | CCT | CCT |
| Revised variables | | | ✓ | ✓ | ✓ | ✓ | ✓ | ✓ | ✓ | ✓ | ✓ | ✓ |
| Four missing districts | | | | | ✓ | ✓ | ✓ | ✓ | ✓ | ✓ | ✓ | ✓ |
| Clustered | | | | | | | ✓ | ✓ | ✓ | ✓ | ✓ | ✓ |
| Regression-specific BW | | | | | | | | | ✓ | ✓ | ✓ | ✓ |
| Bias-corrected | | | | | | | | | | ✓ | ✓ | ✓ |
| Robust RD | | | | | | | | | | | ✓ | ✓ |
| Weights | | | | | | | | | | | | ✓ |

Notes: See notes to Table 3. The regressions reported here differ only in the cutoff.

## C Revised method for estimating general equilibrium effects and elasticities

This appendix reviews and comments on the K23 method for estimating GE effects and elasticities of substitution.

### C.1 Skill premia

K23's starting point for estimating GE effects is this identity, as presented in equation A.20 of the paper's appendix B.5:

$$\log w_{y,D=1} = \ell_{sy,D=1} \log w_{sy,D=1} + \ell_{uy,D=1} \log w_{uy,D=1}$$

where $D$ is a dummy indicator for treatment assignment; $s$ and $u$ index the skilled and unskilled worker subgroups; $w_{y,D=1}$ is the average wage earned by the young in treatment districts; $w_{sy,D=1}$ and $w_{uy,D=1}$ are the same for the skilled and unskilled subgroups; and $\ell_{sy,D=1}$ and $\ell_{uy,D=1}$ are evidently the skilled and unskilled shares in the young labor force of treated districts. The equation is meant to state that a sample average is a weighted average of two subsample averages. As written, it contains two problems. First, earlier in K23 (equation 2), $\ell$ symbols represent absolute quantities of effective labor in a skill-age cell, not fractional shares; they appear to have been redefined here. Second, the equation is formally incorrect because the logarithm of an average is not the average of logarithms. What appears meant is:

$$\overline{\log w_{y,D=1}} = \ell_{sy,D=1} \overline{\log w_{sy,D=1}} + \ell_{uy,D=1} \overline{\log w_{uy,D=1}}$$

where a bar indicates taking the mean over a group. Stipulating this rewriting, K23's key identities for estimating GE effects (K23 equations 17 and 18) may then be developed precisely, as

$$\Delta_D \overline{\log w_y} = \ell_{sy,D=1} \Delta_D \overline{\log w_{sy}} + \ell_{uy,D=1} \Delta_D \overline{\log w_{uy}} + \Delta_D \ell_{sy} \underbrace{\left(\overline{\log w_{sy,D=0}} - \overline{\log w_{uy,D=0}}\right)}_{\beta_{y,D=0}} \tag{1}$$

$$= \ell_{sy,D=0} \Delta_D \overline{\log w_{sy}} + \ell_{uy,D=0} \Delta_D \overline{\log w_{uy}} + \Delta_D \ell_{sy} \underbrace{\left(\overline{\log w_{sy,D=1}} - \overline{\log w_{uy,D=1}}\right)}_{\beta_{y,D=1}} \tag{2}$$

Here, the $\Delta_D$ operator takes differences across the treatment/non-treatment split; e.g., $\Delta_D \overline{\log w_y} = \overline{\log w_{y,D=1}} - \overline{\log w_{y,D=0}}$. As in K23, the expressions labeled $\beta_{y,D=0}$ and $\beta_{y,D=1}$ represent returns to skill within treated or untreated districts.[21] It is readily verified that expanding the definitions on the right side of each equations and cancelling terms produces the left side. In words, the difference in log wages among the young across the treatment/non-treatment split can be decomposed in two similar ways. Both decompose it into a weighted average of differences for the two skills group along with the mathematical effect of movement between the groups.

One could compute $\beta_{y,D=0}$ and $\beta_{y,D=1}$ directly, as implied by the labeled expressions in (1) and (2). Instead, K23 proceeds by first solving for them in (1) and (2):

$$\beta_{y,D=0} = \frac{\Delta_D \overline{\log w_y} - \ell_{sy,D=1} \Delta_D \overline{\log w_{sy}} - \ell_{uy,D=1} \Delta_D \overline{\log w_{uy}}}{\Delta_D \ell_{sy}} \tag{3}$$

[21] K23 subscripts these symbols with $s$. I drop that for clarity. The quantities are differences across the two skill groups, so an index referencing one group, the skilled, is not necessary or meaningful.

$$\beta_{y,D=1} = \frac{\Delta_D \overline{\log w_y} - \ell_{sy,D=0} \Delta_D \overline{\log w_{sy}} - \ell_{uy,D=0} \Delta_D \overline{\log w_{uy}}}{\Delta_D \ell_{sy}} \tag{4}$$

K23 interprets the difference between these two, $\Delta_D \beta_y$, as the general equilibrium effect of DPEP on skilled wages among the young. Though not obvious from the above, the difference is algebraically equivalent to

$$\Delta_D \beta_y = \Delta_D \overline{\log w_{sy}} - \Delta_D \overline{\log w_{uy}} \tag{5}$$

That is: the difference in the skill premium across the treatment/non-treatment divide is a difference in differences.

K23 takes the indirect path represented by (3), (4), and (5) because this allows one to plug in for the $\Delta_D$ terms with FRD estimates, which are described as unbiased. The remaining terms in (3), (4), and (5), such as $\ell_{sy,D=1}$, are computed as subsample means of the relevant variables.

This strategy for estimating returns to skill and GE effects is useful. It also carries several limitations, in principle and in the K23 implementation. One is that most estimates of terms on the right of (3), (4), and (5) contain presumptively endogenous variation. A direct computation of the skill share $\ell_{sy,D=1}$ (the fraction of young workers in program districts who are skilled) depends on DPEP program status ($D$) and on schooling via the skill dummy $s$. Both are taken in K23 as endogenous to the wage outcome. Nor, in the present framework, can a difference such as $\Delta_D \overline{\log w_{sy}}$ be estimated with perfect consistency via FRD, for the samples on each side of the cutoff are restricted to endogenously defined subgroups such as skilled workers.[22]

Moreover, the results from the indirect path to $\beta_{y,D=0}$ and $\beta_{y,D=1}$ are intrinsically less stable than the direct paths implied by the labeling in (1) and (2). The right sides of (3) and (4) are ratios whose denominators are estimated with uncertainty. By the same token, each $\Delta_D$ term, as estimated by FRD, is also prone to large variance: FRD, as an instance of exactly identified instrumental variables estimation has no first or second moments (Kinal 1980).

## C.2 The skill premium

In order to express the skill premia per year of schooling, K23 divides them by an estimate of the average schooling gap between skilled and unskilled workers:

[22] Technically the same argument applies to the splitting of the sample by age in K23 (section V.B) and section 3 above. Districts' population age structure may have been systematically related to their propensity to self-select into DPEP. But age is, if not strictly exogenous, at least predetermined.

$$
\begin{aligned}
\tilde{\beta}_{y,D=0} &= \frac{\beta_{y,D=0}}{\bar{S}_{ys,D=0} - \bar{S}_{yu,D=0}} \\
\tilde{\beta}_{y,D=1} &= \frac{\beta_{y,D=1}}{\bar{S}_{ys,D=0} - \bar{S}_{yu,D=0}} \\
\Delta_D \tilde{\beta}_y &= \frac{\Delta_D \beta_y}{\bar{S}_{ys,D=0} - \bar{S}_{yu,D=0}}
\end{aligned}
\tag{6}
$$

The denominators are the same, a schooling gap in the untreated sample. They too are presumptively endogenous, because of their dependence on $s$ and $u$.

## C.3 Elasticities of substitution

K23 links the empirics to parameters in a hierarchical model of district-level production. The top level of the model is Cobb-Douglas in a district's labor supply ($L_d$) and capital ($K_d$), with constant returns to scale. In turn, $L_d$ is a constant-elasticity-of-substitution (CES) function of the labor supply in various skill cells ($L_{sd}$), which are themselves CES functions of the labor supply in age-skill cells ($\ell_{asd}$).

If each input is paid its marginal product, and if there are two skill groups, indexed by $s$ and $u$, then the skill premium between the two groups within age cohort $a$ and district $d$ is

$$
\log \frac{w_{asd}}{w_{aud}} = \log \frac{\theta_{sd}}{\theta_{ud}} + \left( \frac{1}{\sigma_A} - \frac{1}{\sigma_E} \right) \log \frac{L_{sd}}{L_{ud}} - \frac{1}{\sigma_A} \log \frac{\ell_{asd}}{\ell_{aud}}
$$

where $\sigma_E$ and $\sigma_A$ are the elasticities of substitution in the skill and age levels of the hierarchical model and the $\theta_{sd}$ are the productivity coefficient for $L_{sd}$ in the skill level. If we replace the $d$ indexes with $D = 0$ and $D = 1$, the difference in the skill premium across the treatment/no-treatment split is

$$
\begin{aligned}
\Delta_D \beta_a &:= \log \frac{w_{as,D=1}}{w_{au,D=1}} - \log \frac{w_{as,D=0}}{w_{au,D=0}} \\
&= \underbrace{\Delta_D \log \frac{\theta_{sd}}{\theta_{ud}} + \left( \frac{1}{\sigma_A} - \frac{1}{\sigma_E} \right) \Delta_D \log \frac{L_{sd}}{L_{ud}}}_{\text{independent of age}} - \underbrace{\frac{1}{\sigma_A} \Delta_D \log \frac{\ell_{asd}}{\ell_{aud}}}_{\text{age-specific}}
\end{aligned}
\tag{7}
$$

As noted beneath the equation, only the last term depends on age. K23 appears to assume that it is zero for older workers, i.e., when $a = o$. This amounts to assuming that DPEP did not indirectly affect the schooling levels among the workers too old to have attended DPEP-supported schools (say, via migration or entry into or exit from the workforce):

$$
\Delta_D \log \frac{\ell_{osd}}{\ell_{oud}} = 0 \Rightarrow \frac{\ell_{os,D=1}}{\ell_{ou,D=1}} = \frac{\ell_{os,D=0}}{\ell_{ou,D=0}}
\tag{8}
$$

Plugging in $a = y$ and $a = o$ in (7) then gives

$$\Delta_D\beta_y - \Delta_D\beta_o = -\frac{1}{\sigma_A}\Delta_D \log\frac{\ell_{ysd}}{\ell_{yud}} \tag{9}$$

$$\Delta_D\beta_o = \Delta_D \log\frac{\theta_{sd}}{\theta_{ud}} + \left(\frac{1}{\sigma_A} - \frac{1}{\sigma_E}\right)\Delta_D \log\frac{L_{sd}}{L_{ud}} \tag{10}$$

All quantities in these equations other than the elasticities $\sigma_A$ and $\sigma_E$ are estimated as described after (5). The equations are then solved for the elasticities. K23 estimates $\sigma_A = 5$ and $\sigma_E = 4.24$, the latter using the rough assumption that $\log(\theta_{sd}/\theta_{ud}) = 0$. K23's public data and archive does not precisely document how these values were reached.

## C.4 A revised method for estimating GE effects and elasticities

In light of the conceptual issues described above, and in response to certain implementation details embedded in the K23 code, I propose a revised method for estimating the GE effects and elasticities. The revisions:

1. Those over 75 are excluded from the sample, as in the rest of K23.
2. Those with more than 14 years of schooling are retained, as in the rest of K23.
3. Non-wage-earners are excluded when computing schooling aggregates, as in the rest of K23.
4. Where K23 computes some aggregates as medians, others as means, the revision only takes means, which seems most consistent with the theory.
5. In light of the disadvantages of the indirect route to estimating the skill premia, discussed in section C.1, $\beta_{y,D=0}$ and $\beta_{y,D=1}$ are estimated directly, via the labeled expressions in (1) and (2).
6. This sample is unrepresentative of districts close to the threshold, which are the locus of the FRD estimates. The revision estimates such aggregates by the following procedure, taking $\bar{S}_{ys}$ as an example: First $\Delta\bar{S}_{ys}$ is estimated by FRD. Then $\bar{S}_{ys}$ is computed as the average schooling attainment within the FRD estimation sample, weighting by the FRD kernel. Then the averages in DPEP and non-DPEP districts are estimated as $\bar{S}_{ys,D=0/1} = \bar{S}_{ys} \mp \Delta\bar{S}_{ys}/2$.
7. As shown in equation (6), in order to estimate returns for schooling, K23 divides the treatment skill premium in treatment districts by a schooling difference in non-treatment districts. The revision instead computes $\tilde{\beta}_1 = \beta_1/(\bar{S}_{ys,D=1} - \bar{S}_{yu,D=1})$.
8. The revision drops assumption (8). In particular, successively plugging $a = y$ and $a = o$ into (7) and subtracting the second version from the first gives

$$\Delta_D\beta_y - \Delta_D\beta_o = -\frac{1}{\sigma_A}\left(\Delta_D \log\frac{\ell_{ysd}}{\ell_{yud}} - \Delta_D \log\frac{\ell_{osd}}{\ell_{oud}}\right)$$

The revision estimates $\sigma_A$ by solving for it in this equation rather than in (9).

## C.5 Revised estimates of GE effects and elasticities

Table C–1 compares the results from the K23 methodology with the revised version, the latter performed with and without weights, and with female literacy thresholds of 39.29% and 41%. These estimates correspond to those in the final two columns of Table 5 and Table 6—with all the data and method revisions except possibly weighting. As expected, because the revised methodology minimizes division by quantities estimated with uncertainty, the estimates are much more stable, going by the bootstrap confidence intervals. Of course it may be that, even if the K23 method is not entirely unbiased, it is still less biased than this alternative. At any rate, the overall pattern of results is similar to that seen in section 5. The GE effect estimate is still quite imprecise when using the 39.29% threshold and more precise, but wrong-signed, when using 41%. The estimates of the elasticities of substitution are also imprecise.

**Table C–1. Estimates of general equilibrium effects of DPEP on wages, and of elasticities of substitution, using alternative methodology**

| | Original methodology | New methodology | | | |
|---|---|---|---|---|---|
| Skill premium per year of schooling, non-DPEP districts ($\tilde{\beta}_0$) | 0.199 | 0.115 | 0.261 | 0.055 | 0.051 |
| | (0.00) | (0.00) | (0.04) | (0.00) | (0.00) |
| | [0.11, 0.76] | [0.079, 0.17] | [–0.74, 12.1] | [0.045, 0.069] | [0.035, 0.067] |
| Skill premium per year of schooling, DPEP districts ($\tilde{\beta}_1$) | 0.134 | 0.013 | –0.153 | 0.081 | 0.067 |
| | (0.01) | (0.32) | (0.13) | (0.00) | (0.00) |
| | [0.030, 0.68] | [–0.055, 0.052] | [–9.67, 0.84] | [0.063, 0.093] | [0.047, 0.086] |
| General equilibrium effect ($\Delta_D\tilde{\beta} \coloneqq \tilde{\beta}_1 - \tilde{\beta}_0$) | –0.065 | –0.102 | –0.414 | 0.025 | 0.016 |
| | (0.01) | (0.00) | (0.09) | (0.06) | (0.18) |
| | [–0.19, –0.00] | [–0.23, –0.028] | [–20.1, 2.41] | [–0.0050, 0.047] | [–0.018, 0.049] |
| General equilibrium effect, % ($\Delta_D\tilde{\beta}/\tilde{\beta}_0$) | –32 | –89 | –159 | 46 | 32 |
| | (0.01) | (0.00) | (0.05) | (0.06) | (0.18) |
| | [–77, –3] | [–135, –36] | [–405, 105] | [–7, 102] | [–28, 138] |
| Elasticity of substitution, skill groups ($\sigma_E$) | 4.24 | –6.006 | –11.919 | 3.036 | 5.089 |
| | | (0.28) | (0.59) | (0.36) | (0.45) |
| | | [–60.0, 53.5] | [–455.4, 446.7] | [–36.5, 38.9] | [–41.1, 45.4] |
| Elasticity of substitution, age groups ($\sigma_A$) | 5 | 2.780 | 0.113 | –0.679 | –0.217 |
| | | (0.23) | (0.38) | (0.39) | (0.53) |
| | | [–23.7, 29.5] | [–3.85, 3.09] | [–13.8, 12.7] | [–11.6, 18.2] |
| Female literacy threshold | 39.29% | 39.29% | 39.29% | 41% | 41% |
| Revised variables | | ✓ | ✓ | ✓ | ✓ |
| Four missing districts | | ✓ | ✓ | ✓ | ✓ |
| Clustered | | ✓ | ✓ | ✓ | ✓ |
| Regression-specific BW | | ✓ | ✓ | ✓ | ✓ |
| Bias-corrected | | ✓ | ✓ | ✓ | ✓ |
| Weights | | | ✓ | | ✓ |

*Notes*: "Original" estimate of $\Delta_D\tilde{\beta}$ differs slightly from that in K23 (Table 3), which reports medians of 1,500 bootstrap estimates rather than full-sample estimates. Original values for $\sigma_E$ and $\sigma_A$ are from notes to K23, Table 3. Bootstrap $p$ values are in parentheses, for the one-sided test that a parameter has the sign opposite that estimated. Percentile-based bootstrap 95% confidence intervals are in brackets.